\pdfoutput=1
\documentclass[aps,pra,twocolumn,showpacs,superscriptaddress,a4paper,floatfix]{revtex4}
\usepackage[utf8]{inputenc}
\usepackage{graphicx}
\usepackage{amsmath,amssymb}
\usepackage{hyperref}

\begin{document}

\newcommand {\da} {\ensuremath{d^\dagger}}
\newcommand {\dnn} {\ensuremath{d^{\phantom{\dagger}}}}

\newcommand {\ca} {\ensuremath{c^\dagger}}
\newcommand {\cnn} {\ensuremath{c^{\phantom{\dagger}}}}

\newcommand {\up} {\ensuremath{\uparrow}}
\newcommand {\dn} {\ensuremath{\downarrow}}
\newcommand {\kk} {\ensuremath{{\bf k}}}
\newcommand {\kp} {\ensuremath{{\bf k'}}}
\newcommand {\pp} {\ensuremath{{\bf p}}}
\newcommand {\qq} {\ensuremath{{\bf q}}}
\newcommand {\nbr} {\ensuremath{\langle ij \rangle}}
\newcommand {\myt} {\ensuremath{J}}
\newcommand {\myJ} {\ensuremath{J_{ex}}}
\newcommand {\grfn}{\ensuremath{{\cal G}}}
\newcommand {\dbar}{\ensuremath{{d\!\bar{ }\,}}}

\title{Lifetime of double occupancies in the Fermi-Hubbard model}

\author{Rajdeep Sensarma}
\affiliation{Department of Physics, Harvard University, Cambridge,
Massachusetts 02138, USA}
\affiliation{Condensed Matter Theory Center, Dept. of Physics,
University of Maryland, College Park, Maryland 20742, USA}
\author{David Pekker}
\affiliation{Department of Physics, Harvard University, Cambridge,
Massachusetts 02138, USA}
\author{Ehud Altman}
\affiliation{Department of Condensed Matter Physics, Weizmann Institute,
Rehovot 76100, Israel}
\author{Eugene Demler}
\affiliation{Department of Physics, Harvard University, Cambridge,
Massachusetts 02138, USA}
\author{Niels Strohmaier}
\affiliation{Institute for Quantum Electronics, ETH Z\"{u}rich,
8093 Z\"{u}rich, Switzerland}
\author{Daniel Greif}
\affiliation{Institute for Quantum Electronics, ETH Z\"{u}rich,
8093 Z\"{u}rich, Switzerland}
\author{Robert J\"ordens}
\affiliation{Institute for Quantum Electronics, ETH Z\"{u}rich,
8093 Z\"{u}rich, Switzerland}
\author{Leticia Tarruell}
\affiliation{Institute for Quantum Electronics, ETH Z\"{u}rich,
8093 Z\"{u}rich, Switzerland}
\author{Henning Moritz}
\affiliation{Institute for Quantum Electronics, ETH Z\"{u}rich,
8093 Z\"{u}rich, Switzerland}
\author{Tilman Esslinger}
\affiliation{Institute for Quantum Electronics, ETH Z\"{u}rich,
8093 Z\"{u}rich, Switzerland}

\date{\today}

\begin{abstract}

  We investigate the decay of artificially created double occupancies
  in a repulsive Fermi-Hubbard system in the strongly
  interacting limit using diagrammatic many-body theory and
  experiments with ultracold Fermions on optical lattices. The
  lifetime of the doublons is found to scale exponentially with the ratio
  of the on-site repulsion to the bandwidth. We show that the dominant
  decay process in presence of background holes is the excitation of a
  large number of particle hole pairs to absorb the energy of the
  doublon. We also show that the strongly interacting nature of the
  background state is crucial in obtaining the correct estimate of the
  doublon lifetime in these systems. The theoretical estimates and the
  experimental data are in fair quantitative agreement.

\end{abstract}


\maketitle

The non-equilibrium dynamics of a strongly interacting quantum many-body
system is one of the most complex problems of modern physics. It
encompasses various fields from the cosmology of the early 
universe~\cite{cosmology} or non-equilibrium jet production in high 
energy heavy ion collisions~\cite{rhicjet} to pump-probe experiments
and operation of solid state devices under strong drive~\cite{pump:probe}
in condensed matter physics. There are many
open questions concerning non-equilibrium processes from
both a theoretical and an experimental perspective, especially in the
realm of condensed matter physics.

The theoretical understanding of interacting quantum many body systems
in thermal equilibrium is on a much stronger footing, although
strongly interacting systems like high temperature superconductors
are not yet completely understood. This understanding is based on
paradigms such as the quasiparticle excitations in the Fermi liquid model and
ground states with broken symmetry described in terms of order-parameters
and their fluctuations. The crucial point in all
these paradigms is the hierarchy of energy scales of the
quantum states. By working with a restricted set of states, organized
according to their energy, it is possible to obtain a
simplified model of the system. These low energy descriptions
can capture the response of the system under small perturbations
from equilibrium.  However, in systems far
from equilibrium, there is no organizing principle as the dynamics
couples disparate states with widely different energies and linear
response theory breaks down. This makes it
hard to construct generic paradigms and one needs to solve the full
microscopic Hamiltonian dynamics of an interacting quantum many-body
system.

Some progress has been made for 1D systems, where it is often
possible to obtain exact solutions for the eigenstates of the
Hamiltonian. The absence of thermalization in 1D Bose systems has
been predicted~\cite{bg:notherm, Olshanii2008} and
observed~\cite{bg:notherm:expt} in cold atomic gases. However, these
studies are hard to generalize to higher dimensions.

\begin{figure}[htbp]
  \includegraphics[width=1.\columnwidth,clip=true]{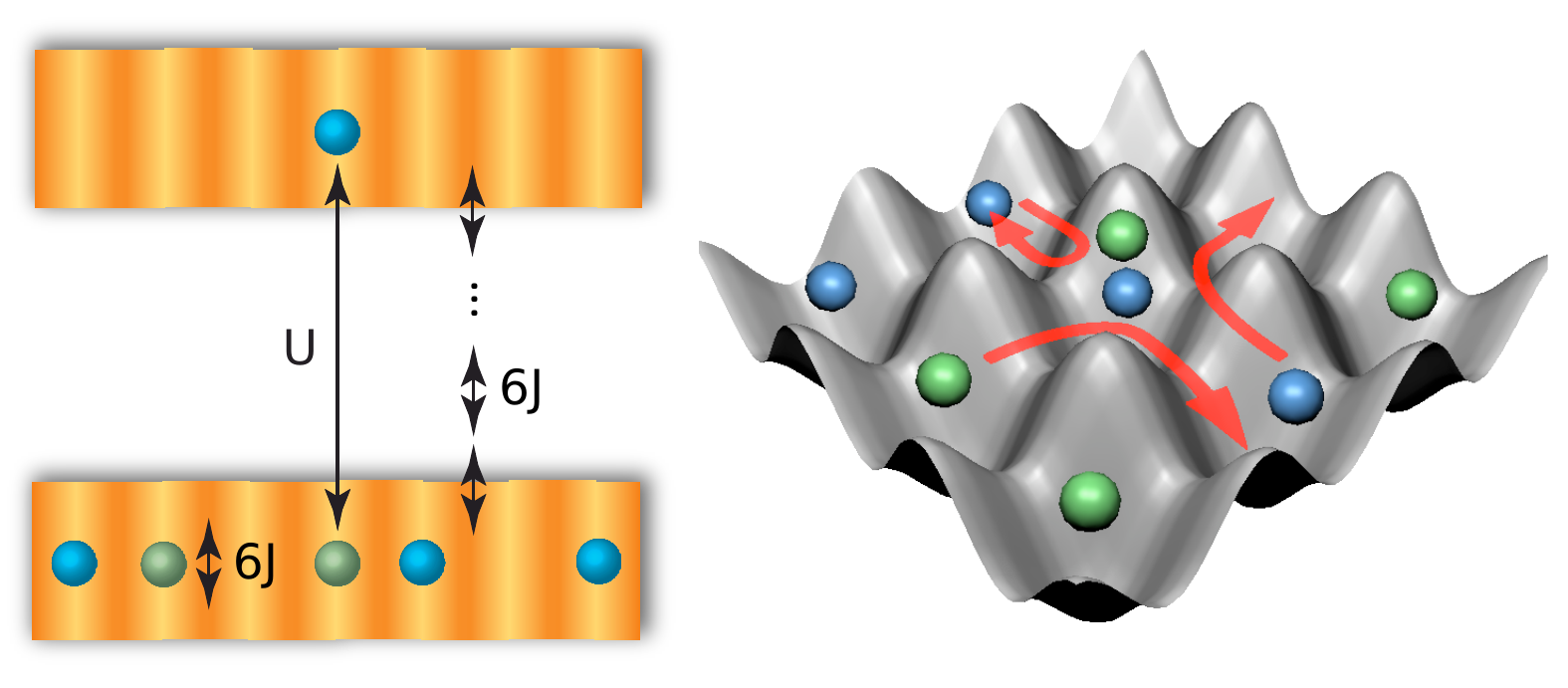}
   \caption{Stability of highly excited states in the single-band Hubbard model. 
   Doubly occupied lattice sites are protected against decay by the on-site 
   interaction energy $U$. The average kinetic energy of a single particle in a 
   periodic potential is half the bandwidth $6J$. Thus the relaxation of 
   excitations requires several scattering partners to maintain energy conservation.}
  \label{fig1:scetch}
\end{figure}

In this context, it is useful to seek answers to concrete and focused
questions involving non-equilibrium dynamics of specific strongly
interacting systems. They have practical importance and help us
gain better understanding of classes of non-equilibrium
processes. Recent advances in controlling ultracold atomic gases with
and without optical lattices have led to their emergence as perfect
systems to study such phenomena.  These systems, which can simulate
strongly interacting model Hamiltonians, are essentially decoupled
from external heat baths and hence the intrinsic non-equilibrium
dynamics of the system can be studied easily. Compared to condensed
matter systems, the low density in these systems results in long
timescales for dynamics. As a result the system can be followed in
real time without the use of ultrafast probes. Further, it is
relatively easy to create and characterize an initial state far from
the ground state, which is crucial since the dynamics depends heavily
on the initial state.

In fact, questions of non-equilibrium dynamics and thermalization
timescales are particularly important for these artificially
engineered strongly correlated systems. Their key feature
is the precise tunability of the Hamiltonian parameters
which has made these systems ideal
for the simulation of strongly interacting many-body Hamiltonians relevant
to condensed matter systems. However, an implicit assumption in this
comparison is that the system is in thermal equilibrium at low
temperatures. In this context it is important to estimate the
thermalization timescales as these systems are always characterized by
a finite sample lifetime. Besides, several proposed methods to prepare
the system in novel phases explicitly depend on adiabatic tuning of
Hamiltonian parameters, which place stronger constraints on the
possible sweep rates than mere demand of thermalization.

An important class of non-equilibrium problems is the decay of a high
energy excitation into low energy excitations. This problem occurs in
diverse contexts like multi-phonon decay of excitons in
semiconductors~\cite{multi:phonon}, pump and probe
experiments~\cite{pump:probe} and dynamics of nuclear
resonances~\cite{nuclear:resonance}.  In this paper, we study 
this problem in the non-equilibrium dynamics of artificially created
double occupancies in the Fermi Hubbard Model in the strongly
interacting regime. Specifically, we will look at the mechanism of
doublon decay in this system and the relation of the doublon lifetime
to the repulsive interaction. We study this dynamics both
experimentally using ultracold Fermions on an optical
lattice~\cite{short:paper} and theoretically using a projected Fermion
model and diagrammatic resummations.

The doublon lifetime has practical implications for the sweep rates of
Hamiltonian parameters in cold atom systems in the following way: The
usual access to the strongly interacting regime is to start
with a weakly interacting system and increase the ratio of interaction
$U$ to the hopping energy $\myt$. As this ratio increases, the
density of doublons in the system in equilibrium should decrease. Thus
the doublon lifetime provides the dominant equilibration timescale for the
system. 
We note here that this problem has structural
similarities with the decay of a deeply bound excitonic state through
multi-phonon processes in semi-conductors~\cite{multi:phonon}, but as we shall see, the
strong Hubbard repulsion modifies the situation in an essential way.

Our main results are (i) The decay of a doublon is a slow process as the
doublon needs to distribute a large energy ($\sim U$) to other
excitations in the system which have a much smaller energy scale. (ii)
The primary mode of decay of the doublon involves creation of
particle-hole pairs in the background system. (iii) The decay rate
scales as $\Gamma\sim C\,\myt \exp (-\alpha \,U/\myt )$
and the decay becomes slower with increasing interaction. We obtain
$C$ and $\alpha$ from experiments and from theoretical
calculation. (iv) We find
that the interactions between pairs of single Fermions, which in our
model are induced by projection, are important and quantitatively
affect the timescale of the decay. Thus the strongly correlated
character of the system changes the dynamics in an essential way.

The paper is organized as follows: In Section I we discuss the various
possible decay mechanisms of the doublon in these systems and give a
scaling argument for the decay rate in each case. In Section II we
describe the experiments and its results. In Section III we discuss
the most relevant decay mechanism in our experiments and develop the
theoretical model for doublon decay. In Section IV we
outline the diagrammatic method to compute the doublon lifetime. In
Section V we discuss the theoretical results and its comparison with
the experiments.  We conclude in section VI by discussing the
importance of these results and future directions. The technical details 
of the theory are described in relevant appendices.

\section{Decay mechanisms for a doublon}

The single-band Hubbard model describing the Fermions on an optical
lattice is given by \cite{Jaksch1998}
\begin{equation} 
H=-\myt\sum_{\nbr\sigma}\ca_{i\sigma}c_{j\sigma}+U\sum_i
n_{i\up}n_{i\dn}.
\end{equation}
At large $U/\myt$, this model has three main energy scales. There is
the energy of double occupancies, given by the Hubbard repulsion $U$,
the kinetic energy of the Fermions given by the tunneling $\myt$ and
the superexchange scale $\myJ=4\myt^2/U$, which governs the spin
dynamics in the system. At large $U/\myt$, these scales are well
separated from each other, $U \gg \myt\gg \myJ$. As we show below,
the separation of the energy scale $U$ from the other energy scales
$\myt$ and $\myJ$ leads to a slow decay of doublons in the system.

In order to decay the doublon has to give up its energy $\sim U$ to
other excitations in the system. Let the typical energy of a possible
excitation be $\epsilon_0$ where $\epsilon_0$ can be either $\sim
\myt$ or $\sim \myJ$ depending on the background state in which the
doublon is propagating. We assume that $\epsilon_0 \ll U$, so that a
large number $n \sim U/\epsilon_0$ of excitations must be created to
satisfy energy constraints. The matrix element for this process can be
calculated by an $n^{th}$ order perturbation theory and is given by
\begin{equation}
M \sim \frac{\myt}{\epsilon_0}\times\frac{\myt}{2\times\epsilon_0}\times\dots\times\frac{\myt}{n\times\epsilon_0}.
\end{equation} 
The decay rate is $\sim M^2$ in units of \myt. Using Stirling's
formula, and the fact that $n\epsilon_0=U$, we 
find that for large $n$ the decay rate scales as 
\begin{equation}
\Gamma \sim \myt (\myt/U)^{\frac{U}{\epsilon_0}} \sim C \myt \exp (-\alpha \, U/\epsilon_0\, \log(U/\myt)) 
\end{equation}
where $C$ and $\alpha$ are constants which we will extract from detailed 
calculations and experimental data. 

In order to discuss the specific decay mechanisms of a doublon, we
need to specify the state of the background system in which the
doublon is propagating. If the system is a homogeneous Mott insulator at
half-filling, the only possible
candidate for transfer of energies are spin excitations with bandwidth
$ \epsilon_0\sim \myJ$. This leads to the decay rate scaling as
$\Gamma \sim \myt \exp (- \alpha \, U^2/\myt^2 \log(U/\myt))$ and is
an extremely slow process. However, if the system is compressible, the
dominant energy transfers are to kinetic energy of the Fermions
through creation of particle-hole pairs with typical energy
$\epsilon_0 \sim \myt$. This leads to the decay rate scaling as
$\Gamma \sim \myt \exp (-\alpha \, U/\myt \log(U/\myt))$. This is a
much faster decay process and will dominate over decay through spin
excitations. We note that compressible states with holes can exist (i)
at the edges of systems with confining traps or (ii) in the bulk of
the system as a result of a large density of doublons created by modulation
spectroscopy. In a trapped system, there is another possibility of
giving up the energy to the potential energy of the Fermions at the
edges. This however involves transfer of particles from the center to 
the edges of the trap and is usually a much slower process for typical 
shallow traps used in cold atom experiments.

As we will see in the next section our experimental system has a lot
of holes and we can eliminate many of the possible decay mechanisms
for our experimental configuration. Therefore, in this Paper we shall
focus on the dominant doublon decay channel involving excitation of
particle hole pairs.

\section{Experiments}

\begin{figure*}[ht]
  \includegraphics[width=\textwidth,clip=true]{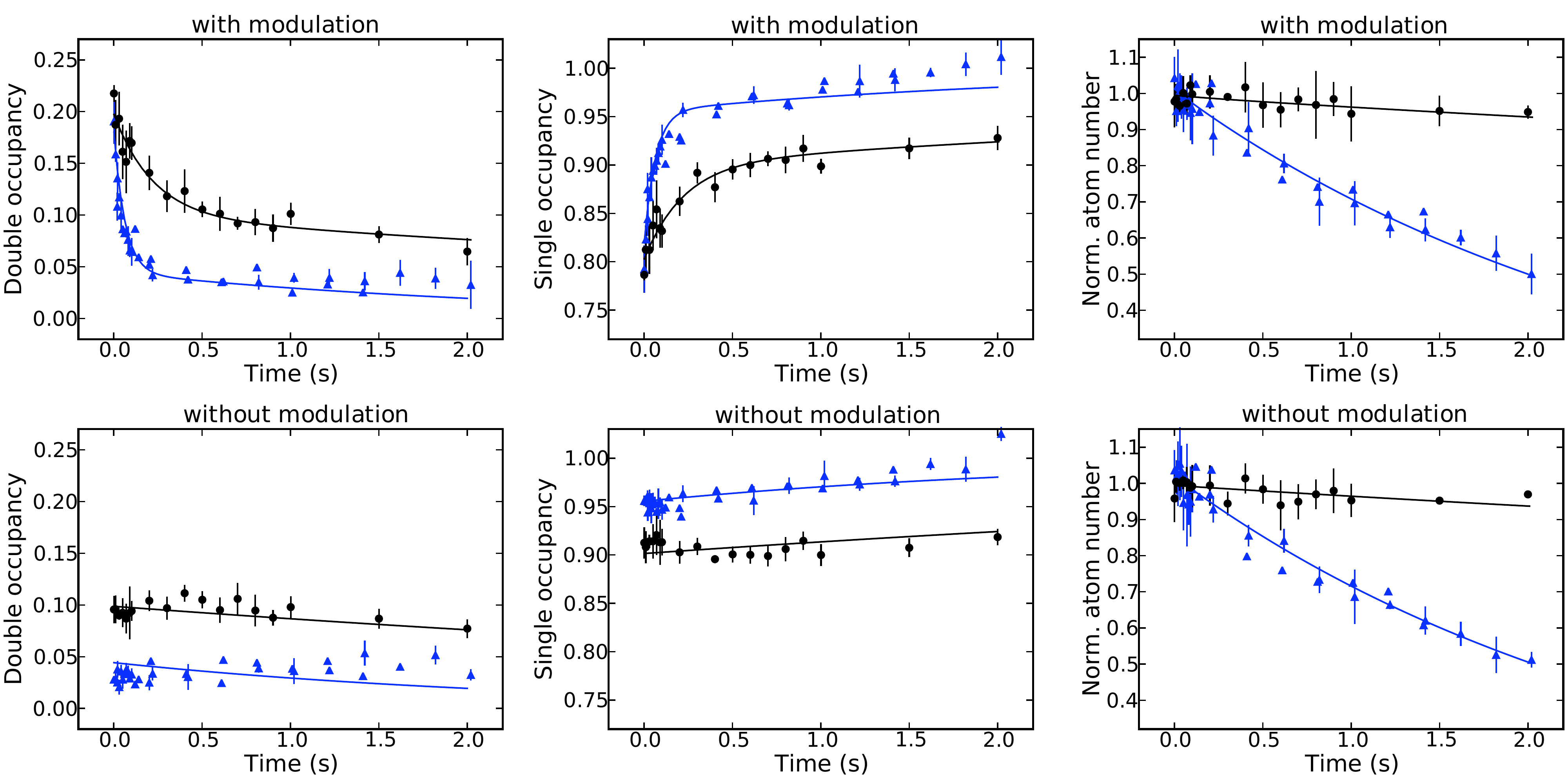}
	
	 \caption{Time evolution of double occupancy, single occupancy and
	 total atom number for different ratios $U/6J$. In the upper row, the
	 system was previously excited via lattice modulation. The bottom row
	 shows the reference measurement for the determination of the residual
	 dynamics. The round data points were recorded using a $m_F = 
	 \left(-9/2,-7/2\right)$ spin mixture with $U/h = 1.4\,\text{kHz}$ 
	 and $J/h = 70\,\text{Hz}$, whereas the triangular data points show a
	 $\left(-9/2,-5/2\right)$ mixture with $U/h = 3.2\,\text{kHz}$ and
	 $J/h = 100\,\text{Hz}$. The solid lines are simultaneous fits of the
	 integrated population equations of Eq.~\ref{eqn:popdecay}. The total atom
	 numbers are scaled to the initial values. Single occupancy and double
	 occupancy are the fraction of atoms residing on sites of the
	 respective type. Due to different detection
	 efficiencies for hyperfine states the sum of double and single
	 occupancy can be higher than one. Error bars denote the statistical
	 error of at least four identical measurements.}

  \label{rawdata59}
\end{figure*}

This section describes the experimental steps towards the observation of
doublon relaxation: Initially, a sample of repulsively interacting,
ultracold fermions is produced and loaded into an optical lattice.
Starting from this equilibrium state, we create additional double
occupancies via lattice modulation. Immediately after this excitation we
measure the time evolution of the double occupancy. We remove the
influence of inelastic loss processes by comparing to a reference
measurement and extract the elastic doublon lifetime using a simple rate
equations model. Finally, this elastic lifetime is normalized with the
tunneling time $J/h$ and found to depend exponentially on $U/6J$.

The experimental sequence used to produce quantum degenerate Fermi
gases has been described in detail in previous work
\cite{Jordens2008}. In brief, we prepare $(50\pm 5)\times10^3$
$^{40}$K atoms at temperatures below $15\%$ of the Fermi temperature
$T_{\mathrm{F}}$ in a balanced mixture of two magnetic sublevels of
the $F=9/2$ hyperfine manifold. The confinement is given by a crossed
beam dipole trap with trapping frequencies $\omega_{x,y,z} = 
2\pi\times(35, 23, 120)\,\text{Hz}$. To access a wide range of
repulsive interactions we make use of two magnetic Feshbach
resonances. With a $m_F=\left(-9/2,-7/2\right)$ spin
mixture, we realize scattering lengths of $98\,a_0$ and $131\,a_0$,
where $a_0$ is the Bohr radius \cite{Regal2003}. The
$\left(-9/2,-5/2\right)$ spin mixture allows us to
reach the strongly repulsive regime with scattering lengths of $374\,a_0$, $571\,a_0$ and $672\,a_0$ \cite{newResonance}. After adjusting
the scattering length to the desired value, we add a three-dimensional
optical lattice of simple cubic symmetry. The lattice depth is
increased in $200\,\text{ms}$ to final values between $6.5 E_R$
and $12.5 E_R$ in units of the recoil energy
$E_R=h^2/2m\lambda^2$. Here $h$ is Planck's constant,
$m$ the atomic mass and $\lambda=1064\,\text{nm}$ the wavelength of
the lattice beams.  The lattice beams have Gaussian profiles with
$1/e^2$ radii of $w_{x,y,z}= (160, 180, 160)\,\mu\text{m}$ at the
position of the atoms. For a given scattering length and lattice
depth, $J$ and $U$ are inferred from Wannier functions
\cite{Jordens2008,Jaksch1998}. Their statistical and systematic errors
are dominated by the lattice calibration and the accuracies in width
and position of the two Feshbach resonances \cite{Regal2003,
newResonance}.

Depending on $U$ and $J$ the final states of the system range from
metallic to Mott insulating phases, but always with a double occupancy
below $15\%$. This equilibrium system is now excited by a sinusoidal
modulation of the lattice depth with a frequency close to $U/h$. This causes an increase of the double
occupancy between 5 and $20\%$ as compared to the initial state. The
modulation amplitude is $10\%$ on all three axes, while the modulation
duration was chosen such that the amount of doubly occupied lattice
sites saturates \cite{Kollath2006,Hassler2008,Jordens2008, Huber2008,
Sensarma2009}. The system is now in a highly excited non-equilibrium
state with double occupancies between 15 and $35\%$.

After free evolution at the initial lattice depth and interaction strength for a variable hold time up 
to $4$\,s we probe the remaining double occupancy of the system. This is accomplished by a sudden increase of the lattice depth to 
$30\,E_{\mathrm{R}}$, which prevents further tunneling. We then measure the 
amount of atoms residing on singly (doubly) occupied sites $N_{\mathrm{s}}$ 
($N_{\mathrm{d}}$) by encoding the double occupancy into a previously 
unpopulated spin state using radio frequency spectroscopy \cite{Jordens2008}. Combining 
Stern-Gerlach separation and absorption imaging we obtain the single occupancy 
$n_{\mathrm{s}}=N_{\mathrm{s}}/N_{\mathrm{tot}}$, double occupancy  
$n_{\mathrm{d}}=N_{\mathrm{d}}/N_{\mathrm{tot}}$ and total atom number 
$N_{\mathrm{tot}}=N_{\mathrm{s}}+N_{\mathrm{d}}$.

The time evolution of the double and single occupancy and of the total
atom number is shown for two different parameter sets in the upper row
of Fig.~\ref{rawdata59}. In both cases, the double occupancy decays
exponentially within the observation time, and the single occupancy
rises accordingly. The time evolution of the total atom number,
however, exhibits a remarkable difference between the
$\left(m_{\mathrm{F}}=-9/2,-7/2\right)$ and the $\left(m_{\mathrm{F}}
  = -9/2,-5/2\right)$ spin mixture: Whilst the atom number of the
$\left( -9/2,-7/2\right)$ sample remains rather constant, the $\left(
  -9/2,-5/2\right)$ sample suffers from an atom number reduction of
50\,$\%$ within $2$\,s. This behavior can be observed for all parameter
sets and is a consequence of the shorter lifetime of a $\left(
  -9/2,-5/2\right)$ spin mixture.

The only relevant process described by the Fermi-Hubbard model is the
decay of a doublon into two single particles which remain within the
system. The time associated with this process will be called doublon
lifetime. In an experiment, inelastic processes may occur, resulting
in atoms exiting the system. For a valid comparison with theory it is
therefore crucial that these processes do not interfere with the
determination of the doublon lifetime.  In the following, we show how
we eliminate the influence of inelastic loss processes on the
observation of the doublon decay.

For every dataset on doublon decay after lattice modulation, we record
a corresponding reference dataset without lattice modulation, but with
the same system parameters. Two of these reference datasets are
presented in the bottom row of Fig.~\ref{rawdata59}. They show the
dynamics of double occupancies and atom number governed by inelastic
processes, which are not taken into account by the Fermi-Hubbard
model.

Combining these two measurements, we can unambiguously extract the
doublon lifetime by simultaneously fitting a system of coupled rate
equations. They describe the population dynamics in the optical
lattice, considering three general processes:

\begin{eqnarray}
\label{eqn:popdecay}
\Delta\dot{N}_{\mathrm{d}}&=&-\left(\frac{1}{\tau_{\mathrm{D}}}+\frac{1}
{\tau_{\mathrm{in}}}+\frac{1}{\tau_{\mathrm{loss}}}\right)\Delta 
N_{\mathrm{d}}\nonumber\\
\dot{N}_{\mathrm{d},0}&=&-\left(\frac{1}{\tau_{\mathrm{in}}}+\frac{1}
{\tau_{\mathrm{loss}}}\right)N_{\mathrm{d},0}\\
\dot{N}_{\mathrm{s}}&=&\frac{1}{\tau_{\mathrm{D}}}\Delta N_{\mathrm{d}} - 
\frac{1}{\tau_{\mathrm{loss}}}N_{\mathrm{s}}\nonumber
\end{eqnarray}
The total number of atoms on doubly occupied sites $N_{\mathrm{d}}$ is
written as the sum of the equilibrium population $N_{\mathrm{d},0}$
and the additional amount of double occupancy $\Delta N_{\mathrm{d}}$
created by the lattice modulation. The three time constants correspond
to three independent local decay processes differing in the type of
site they affect: $\tau_{\mathrm{D}}$ describes the population flow
from doubly occupied to singly occupied lattice sites visible as a
decay of double occupancy within $0.01-1\,\text{s}$ that is accompanied
by a rise of the single occupancy. We
identify this time with the lifetime of doublons. The other two times
denote loss time constants, which lead to a reduction of the total
atom number: $\tau_{\mathrm{loss}}$ corresponds to losses affecting
both site types in the same manner, which is only observed in the
total atom number. Additional inelastic losses on doubly occupied
sites are summarized by $\tau_{\mathrm{in}}$, visible as a
simultaneous decay of both the total atom number and double occupancy.
This model does not account for changes of the decay times during the
decay or for higher order terms in the rate equations.

Since the modulation has no influence on the evolution of the total
atom number, this procedure removes the influence of $\tau_\text{in}$
and $\tau_\text{loss}$. A reliable determination of the doublon
lifetime $\tau_\text{D}$ is thus possible if it differs significantly
from the loss times. The model and the observation are found to agree
very well within experimental uncertainties, as can be seen in
Fig.~\ref{rawdata59}.

\begin{figure}[htb]
  \includegraphics[width=0.71\columnwidth,clip=true]{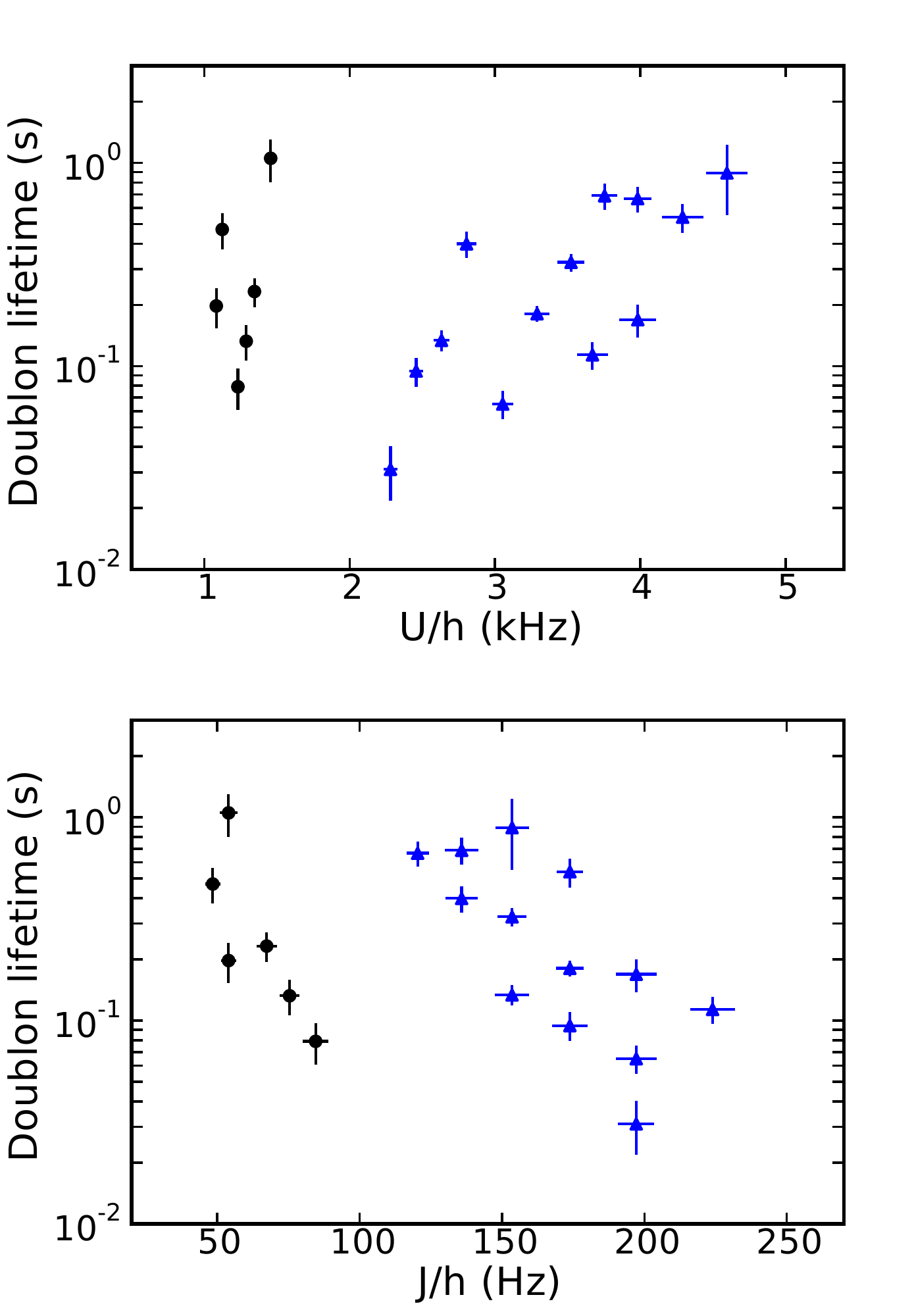}

		\caption{Doublon lifetime as a function of $U$ and $J$. The round
		data points show the fit results to datasets as shown
		in Fig~.~\ref{rawdata59}, obtained with a $\left(-9/2,-7/2\right)$
		spin mixture while the triangular points correspond to the
		$\left(-9/2, -5/2\right)$ mixture. Error bars denote the
		confidence intervals of the lifetime fits and the statistical errors
		in $U$ and $J$.}

  \label{lifetimeUJ}
\end{figure}

We measure this doublon lifetime for various tunneling and interaction
strengths, covering a parameter range where $J$ and $U$ each differ by
at least a factor of four. The determined lifetimes vary over two
orders of magnitude, as shown in Fig.~\ref{lifetimeUJ}. Furthermore,
the lifetime clearly does not depend on the tunneling energy or the
interaction energy alone.

Since the tunneling time $h/J$ is the dominant timescale of dynamics
in an optical lattice, it appears natural to express the doublon
lifetime in units of $h/J$. After this rescaling, we found that, to a
good extent, the doublon lifetime only depends on the ratio $U/6J$.

\begin{figure}[b]
  \includegraphics[width=\columnwidth,clip=true]{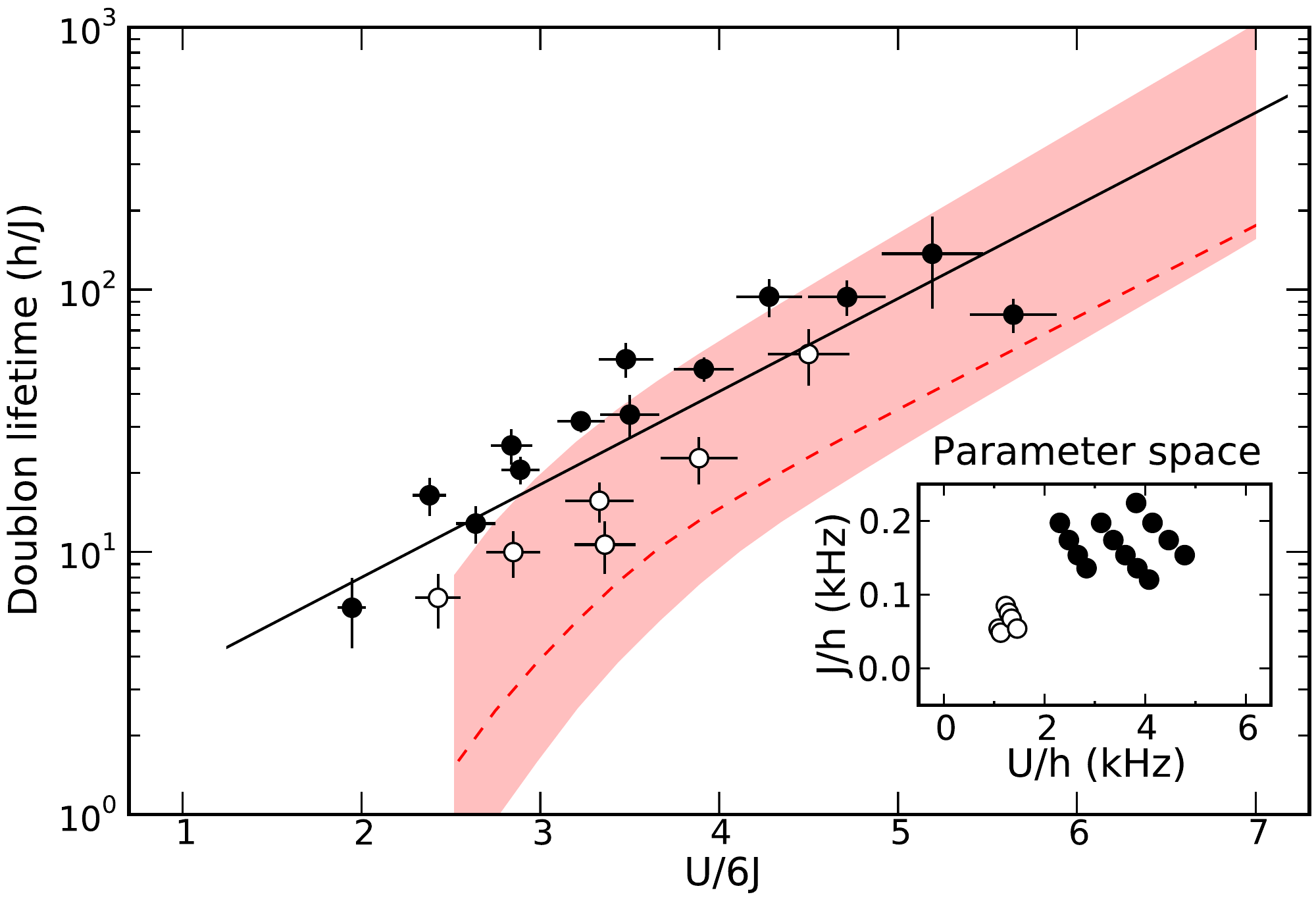}

	 \caption{Semilogarithmic plot of doublon lifetime $\tau_D$ vs.
	 $U/6J$. The lifetime is extracted from datasets as shown in
	 Fig.~\ref{rawdata59}. Solid and hollow circles denote the
	 $\left(-9/2,-5/2\right)$ and $\left(-9/2,-7/2\right)$ spin mixture
	 respectively, while the dashed line shows the theoretical result at
	 half filling. The solid line is a fit of Eq.~\ref{scalingeqn} to the
	 experimental data, yielding $\alpha = 0.82\pm 0.08$, whereas for the
	 theory curve the asymptotic slope at large $U/6J$ is $\alpha_T =
	 0.80$.  The shaded corridor was obtained by varying the filling
	 factor in the calculation by $0.3$. This has only a weak effect on
	 the slope. The inset shows the parameters used to realize the
	 different values of $U/6J$. Error bars denote the confidence
	 intervals of the lifetime fits and the statistical errors in $U/6J$.
	 The systematic errors in $U/6J$ and $\tau_D =h/J$ are estimated to be
	 30\% and 25\%, respectively.}
 
  \label{fig3:scalingdata}
\end{figure}

Fig.~\ref{fig3:scalingdata} shows the doublon lifetime in units of the
tunneling time versus $U/6J$ on a logarithmic scale. Remarkably, over
the entire parameter range the data collapses in a corridor and can be described by an exponential function of the form:
\begin{equation}
\label{scalingeqn}
\frac{\tau_D}{h/J} =C\, \exp\left(\alpha\frac{U}{6J}\right).
\end{equation}
The scaling exponent $\alpha$ is found to be $\alpha=0.82\pm0.08$ with
$C=1.6\pm 0.9$. This is in reasonable quantitative agreement with the
following calculation of the doublon lifetime.

The slight offset between the two spin mixtures in Fig.~\ref{fig3:scalingdata} could be due to the fact that the absolute values for $U$ and $J$ differ significantly between the $\left(-9/2,-5/2\right)$ and the $\left(-9/2,-7/2\right)$ mixture \cite{offsetalpha}. Whilst the ratio between interaction energy and kinetic energy $U/6J$, which dominates the dynamics, lies in the same range, the absolute values also matter in an inhomogeneous system. For the $\left(-9/2,-7/2\right)$ mixture the higher ratio of chemical potential to on-site interaction is expected to lead to a higher filling in the trap centre and consequently to a higher equilibrium double occupancy $N_{d,0}$ than for the $\left(-9/2,-5/2\right)$ mixture. It is conceivable that this difference modifies the dynamics of doublon creation and doublon relaxation.

In additional measurements we examined the dependence of the doublon
lifetime on the initial double occupancy $\Delta N_d$ and on the total
atom number $N$. In the former case, we reduced the lattice modulation
amplitude from 10\% to 5\%, resulting in $\Delta N_d = 9\%$ instead of
$\Delta N_d = 18\%$, while keeping all other parameters constant with
$U/6J=4.5$. The measured lifetimes agree within the error bars, they are
$\tau_{\mathrm{D},5\%}=(77\pm 25)\times h/J$ and $\tau_{\mathrm{D},10\%}
= (58 \pm 10)\times h/J$, respectively. In the latter case, we prepared
two otherwise identical samples at $U/6J=3.4$ with $N=(49\pm7)\times
10^3$ atoms and with $N=(26\pm4)\times 10^3$ atoms, respectively,
yielding $\tau_{\mathrm{D},49000} = (11 \pm 2)\times h/J$ and
$\tau_{\mathrm{D},26000} = (19 \pm 2)\times h/J$.

This shows that, although there is a dependence on the total density and on the doublon density, these effects are small compared to the dominant scaling with $U/6J$. Their systematic study is beyond the scope of this work. 

\section{Theoretical Model of Doublon Decay}

We consider the decay of an isolated doublon moving in the homogeneous
background of a compressible state of single Fermions. Before
constructing a model for doublon decay, we focus on the dominant
mechanism of decay. In the experiments, lattice modulation created
$15-35 \% $ double occupancies. Assuming an initial half-filled
system, half the amount of holes were also created in the system. At
these hole densities, the kinetic energy assisted decay scaling
as $\sim \exp(-U/\myt)$ is much faster than the spin fluctuation or
doublon-doublon collision assisted decay which scale as $\sim
\exp(-U^2/\myt^2)$ \cite{Mott:decay}.  Further, the population of
higher bands can be excluded, since $U$ is always smaller than half
the band gap. We also note that as the difference between $U$ and the
chemical potential is always positive, confinement assisted decay of
doublons near the edge of the cloud is unlikely, as the accessible
confinement energy is not very large, and the tunneling rate is very
small. Finally, a homogeneous compressible background is justified
since most of the doublons are created in the central region of the
trap, where the filling is highest, and decay at most within a few
sites of where they are produced The estimated travel distance for a
random walk during the decay process is not more than
$\sqrt{\tau_{\mathrm{D}} J/h} \lesssim 10$ sites, which is less than
the cloud radius.

In our experiments, the doublons and holes are created at finite
density by driving the system with optical lattice modulations. The
relaxation of the system to equilibrium involves two very different
time scales. The first timescale is associated with the relaxation of
holes and doublons to a state of quasi-equilibrium without the decay of
doublons. The second timescale, which is the focus of this paper, is
associated with the decay of doublons into singles. We expect that the
second timescale is much slower than the first. Moreover, we expect
that non-linear effects of doublon decay as doublon-doublon scattering
can be neglected since their kinetic energy $\sim \myt^2/U$
is small. Thus in this paper we consider the
problem of the decay of a single doublon in the background of
equilibrated Fermions.

To construct our model Hamiltonian, we explicitly treat the doublon
as a separate entity from the background Fermions. This approximation
is justified in the strongly interacting limit due to the separation
of doublon and background Fermion time-scales. We split the complete
Hamiltonian of the system into three parts
\begin{align}
H=H_f+H_d+H_{fd},
\end{align}
where $H_f$ describes the background Fermion subsystem---which we
model as the projected Fermi sea, $H_d$ describes the on-site
interaction of the pair of Fermions that make up a doublon, and
$H_{fd}$ describes the Fermion-doublon interaction. The details of how
to separate the Fermi-Hubbard Hamiltonian into the above three parts
via projection operators are discussed in Appendix ~\ref{appendix:Model}. The projection operators induce
interactions in the Fermion subsystem as well as between the Fermions
and the doublons. The Fermion doublon interactions are responsible 
for the doublon decay, and the Fermion-Fermion interactions modify the 
lifetime substantially.

As mentioned above, we expect hole density in these systems to be $\sim
15\%$. At such high hole
densities the projected Fermi sea is a good approximation for the
background state. Further the temperature of the system is high enough
($T\sim \myt$) to prevent formation of more ordered states like
superfluids.

Except for the single doublon that is undergoing decay, the large
energy cost of double occupancies is taken into account by projecting
out configurations with double occupancies from a simple Fermi sea. In
the projected subspace, the Fermions can only hop in the presence of
empty sites (holes) and are governed by the effective Hamiltonian
\begin{equation}
H_f=-\myt \sum_{\nbr,\sigma}(1-n_{i\overline{\sigma}})\ca_{i\sigma} \cnn_{j\sigma}(1-n_{j\overline{\sigma}})
-\mu\sum_{i,\sigma} \ca_{i\sigma} \cnn_{i\sigma}
\label{Eq:Hf1}
\end{equation}
where $\ca_{i\sigma}$ creates a Fermion with spin $\sigma$,
$n_{i\sigma}$ is the corresponding number operator, and
$\mu$ is the chemical potential. Expanding out the
Hamiltonian one gets $H_f=H_f^0+H_p$, with
\begin{eqnarray}
H_f^0 &= &-\myt \sum_{\nbr,\sigma}\ca_{i\sigma} \cnn_{j\sigma} 
-\mu\sum_{i,\sigma} \ca_{i\sigma} \cnn_{i\sigma},
\label{Eq:HF}\\
H_p&=&\myt_1\sum_{\nbr,\sigma}n_{i\overline{\sigma}} \ca_{i\sigma}
\cnn_{j\sigma}+\ca_{i\sigma} \cnn_{j\sigma}n_{j\overline{\sigma}},
\label{Eq:H_p}
\end{eqnarray}
where we have replaced $\myt$ by $\myt_1$ in the second term. $\myt_1$
will be treated as a perturbation parameter to organize the
calculation but we will put $\myt_1=\myt$ at the end of the
calculation. $H_p$, coming from the projection operators can thus be
interpreted as a Fermion-Fermion scattering term which leads to the
creation of particle-hole pairs. We thus see that projection induces
interaction between the Fermions. 

We note that the scattering is always between Fermions of opposite
spins. Since we will be interested in calculating Feynman diagrams, we
note that the interaction vertex for the Fermion Fermion scattering
can be written as $\myt_1(\gamma_\kk+\gamma_{\kk-\qq})$, where $\kk$
and $\kk-\qq$ are momenta of the incoming and outgoing Fermion with
the same spin and $\gamma_\kk=2 \sum_{i=1}^D \cos k_i$ in $D$
dimensions. This is depicted in first row of Table ~\ref{Table:Rules}.

Throughout our treatment, we leave out terms such as
$-\myt\, n_{i\overline{\sigma}}\ca_{i\sigma}
\cnn_{j\sigma} n_{j\overline{\sigma}}$ in Eq.~(\ref{Eq:Hf1})
involving six or more fermion creation/annihilation operators.
Intuitively such terms are rare because they
involve collisions of multiple particles.

\begin{table}
\begin{tabular}{|c|c|c|}
\hline
Interaction &  Diagram  & 
$
\begin{array}{c}
\text{Vertex}\\
\left[
\begin{array}{c}
\text{Momentum} \\
\text{Avg. Vertex}
\end{array}
\right]\end{array}$ \\
\hline
\parbox{3.0cm}{Fermion-Fermion\\ Scattering} &\includegraphics[scale=0.4]{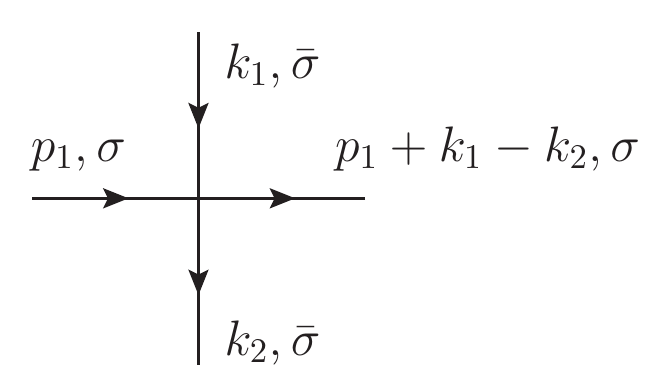} & 
\parbox{2.5cm}{
$
\begin{array}{c}
J (\gamma_{\kk_1}+\gamma_{\kk_2}) \\
 \\
\left[J \sqrt{2 z} \, \right]
\end{array}
$
}
\\
\hline\parbox{3.0cm}{
Doublon-Fermion\\ Scattering} &\includegraphics[scale=0.4]{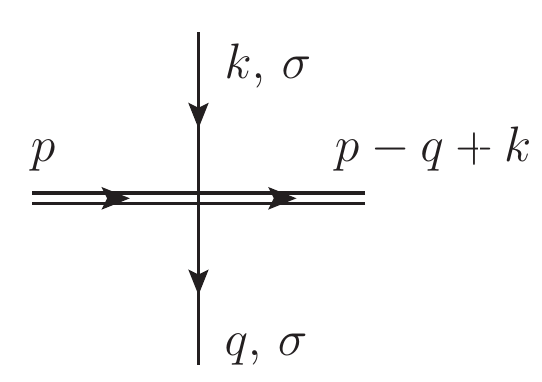} & 
$\begin{array}{c}
J (\gamma_{\pp-\qq}+\gamma_\pp+\gamma_\qq)\\
 \\
\left[J \sqrt{z} \, \right]
\end{array}$ \\
\hline
Doublon Decay &\includegraphics[scale=0.4]{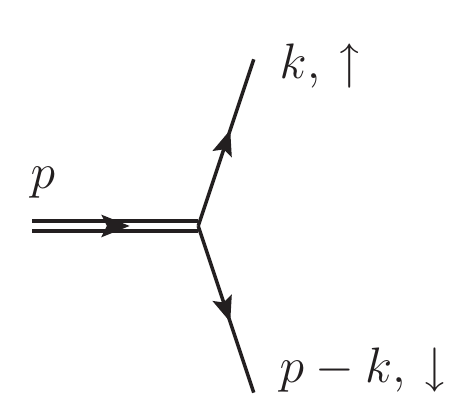} & 
$ \begin{array}{c}
J  (\gamma_{\kk}+\gamma_{\pp-\kk})\\
\\
\left[J \sqrt{2 z}\, \right]
\end{array}$ \\
\hline
\end{tabular}
\caption{
  Interaction vertices for different processes in the model for doublon decay. 
  The single lines are Fermion propagators while the double lines are doublon 
  propagators. The top entry in the right-most column is the corresponding 
  vertex function, while the bottom entry is the ``momentum averaged'' vertex 
  function that we use in the our resummation technique. The first row 
  corresponds to $H_p$ (Eq.~\ref{Eq:H_p}) while the next two rows correspond
  to the two terms that make up $H_{fd}$ (Eq.~\ref{Eq:H_fd}). Here 
  $\gamma_k = 2 (\cos(k_x) + \cos(k_y) + \cos(k_z))$ and $z$ is the coordination number,
  $z=6$ for 3D cubic lattice.}
\label{Table:Rules}
\end{table}

We now consider the decay of a single doublon in this background
state. The doublon ($d$) and Fermion-doublon ($fd$) Hamiltonians can
be written as
\begin{align}
H_{d}&=U\sum_i \da_id_i, \\
\label{Eq:H_fd}
H_{fd}&= \myt \sum_{\nbr\sigma} (\da_i \dnn_i+\da_j \dnn_j+\da_j \dnn_i) \ca_{i\sigma}\cnn_{j\sigma} \\
&\nonumber \quad\quad\quad\quad\quad + d_i\sigma \ca_{i\sigma}\ca_{j\overline{\sigma}}(1-n_{j\sigma})+ \text{H.c.},
\end{align}
where $\text{H.c.}$ stands for the Hermitian conjugate of the preceding
term.  The doublon interacts
with the Fermions in two different ways: (i) it can scatter off a
Fermion leading to the hopping of doublons with back-flow of Fermions;
(ii) it can decay by creating a singlet particle-particle pair.

The interaction vertices of the doublon with the Fermions are given in
second and third rows of Table ~\ref{Table:Rules}. The vertex for
scattering off particle-hole pairs is $\myt
(\gamma_{\pp-\qq}+\gamma_\qq+\gamma_\kk)$, where $\pp$ is the momentum
of the incoming doublon, $\qq$ is the momentum of the outgoing Fermion
and $\kk$ the momentum of the outgoing hole. The corresponding vertex
for decay through singlet creation is given by $\myt (\gamma_\kk+\gamma_{\pp-\kk})$ 
where $\kk$ and $\pp-\kk$ are the momenta of the Fermions created.

We assume that we are looking at the decay of a single doublon
i.e. while the doublon is affected by the presence of the background
Fermions, the Fermions are unaffected by the presence of the
doublon. The motivation for this assumption is that the experimentally
observed decay rate depends only weakly on the doublon density.

\section{Diagrammatic Computation of Doublon Lifetime}
\label{Sec:diagramatics}
Our strategy for finding the lifetime of a doublon is to calculate the 
doublon Green function 
\begin{equation}
\grfn_d(\omega)=[\omega -U -\Sigma_d(\omega)]^{-1},
\label{Eq:dbl_gfn}
\end{equation}
where $\Sigma_d$ is the self-energy arising from interaction with
Fermions.  The imaginary part of the self-energy at $\omega =U$ then
gives the decay rate $\Gamma$ and its inverse is the required lifetime
$\tau$. Since we are interested in the high frequency response, the
momentum dependence of the self energy should be negligible in this
limit. 

We perform the calculation at $T=0$, where the relation between
imaginary part of the self-energy and decay rate is exact. At finite
temperatures $\text{Im} ~ \Sigma(\omega)$ has an extra contribution
due to scattering on particle-hole pairs created by thermal
fluctuations. Thus, we must compute the scattering rate separately,
and subtract it from $\text{Im} ~ \Sigma(\omega)$ to obtain the decay
rate.  However, since we are looking at frequencies $\sim U$, ignoring
thermal fluctuations is justified for $T \ll U$, which is the regime
of interest.

Physically, there are two important processes for the doublon decay.
A doublon can lose its energy either by creating a large number of
particle-hole pairs, each with an energy $\sim \myt$, or by creating a
few high energy particle-hole pairs, each of which is unstable and
creates a shower of particle hole pairs of low energies. The first
process is a high order diagram in the doublon self-energy while the
second process comes from high order diagrams in the Fermion
self-energy. We find that combinations of both processes give
important contributions to the doublon decay rate. 

\begin{figure}
\includegraphics[scale=1]{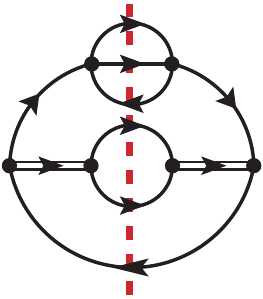}
\caption{A Typical doublon self-energy diagram. The double lines are bare 
doublon propagators while the single lines are bare Fermion propagators. The
dashed line cuts the diagram in half and shows the final products of the 
process represented by this diagram}
\label{Fig:cutdiagram}
\end{figure}

The typical doublon self-energy diagram (Fig.~\ref{Fig:cutdiagram})
depicts a process of creation of a number of particle and hole
excitations. Since we are interested in the imaginary part of the self
energy at $\omega=U$, the Fermion lines crossing the dashed line which
cuts the diagram in half should be on-shell and their energies must
add up to $U$. The leading order contributions to the decay rate thus
come from the diagrams which maximize the number of Fermions that
cross the dashed line while minimizing the number of interaction
vertices.

Our approach for obtaining the doublon self energy consists of
(1)~obtaining the projected Fermi sea Green function, and (2)~using it
to obtain the doublon self-energy.  We make the dilute doublon
approximation, and assume that the Fermion Green function is
independent of the doublon Green function. We proceed by formulating a
diagrammatic resummation technique for the doublon self-energy in the
following subsection. In doing so, we relate the doublon self-energy
to the Fermion Green function, which we calculate in the next two
subsections.

\subsection{Doublon Self-Energy}
\label{Subsection:DoublonSelfEnergy}
\begin{figure}
\includegraphics[scale=0.6]{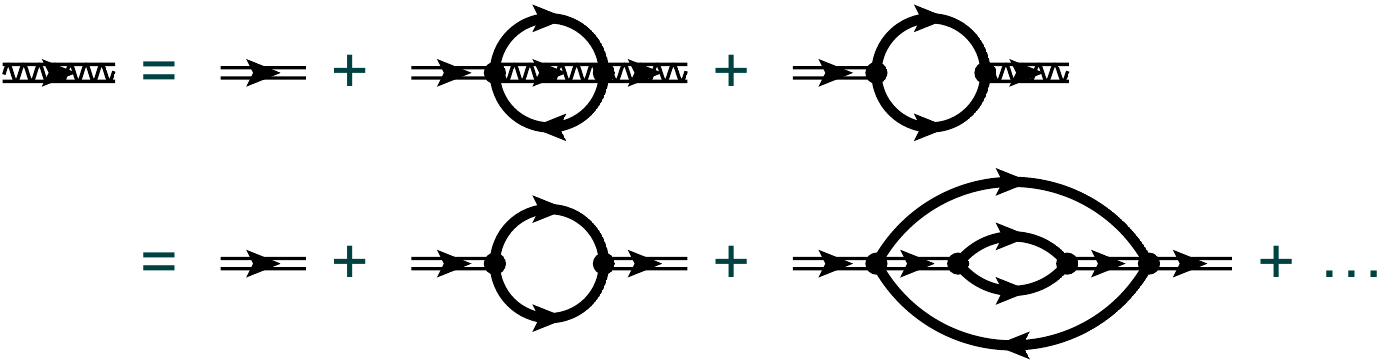}
\caption{Self-consistency equation for the doublon propagator (top)
  and some typical diagrams that make up the full propagator
  (bottom). Thin double-lines indicate the bare doublon propagator, the
  double-lines with a squiggle the full (resummed) doublon propagator,
  and thick single-lines the full (resummed) Fermion propagator.}
\label{Fig:DoublonGF}
\end{figure}

For large $U/\myt$, doublon decays into a large number of
particle-hole pairs, and therefore one needs to compute high order
diagrams to obtain the doublon self-energy (for creation of $n$ pairs,
one needs to compute $\sim 2n !$  diagrams). It is then much
preferable to resum a class of diagrams, rather than evaluate an
exponentially increasing number of them. We use a self-consistent
non-crossing approximation to achieve this resummation.  The
propagator diagrams are shown in Fig.~\ref{Fig:DoublonGF}, where the
doublon lines with squiggles represent the full doublon Green function
to be obtained self-consistently and the thick single lines are the
Fermion propagators.

At this point, we make an additional approximation, and replace the
$\kk$-dependent vertex functions $\Lambda_\kk$ by momentum averaged vertex
functions $\sqrt{\langle \Lambda_\kk^2 \rangle}$ listed in
Table~\ref{Table:Rules}. The basis of this approximation, is that
within our resummation scheme, the vertex functions always occur in
pairs with identical and largely arbitrary momentum indices, as can
be seen from the self-consistent equation represented in
Fig.~\ref{Fig:DoublonGF}.  The self-consistent equation, therefore,
contains the product of this pair of vertex functions, and we replace
this product by its momentum averaged value.

Having replaced the momentum-dependent vertex functions by
momentum-independent ones, we can replace Green functions and
self-energies by their momentum averaged counterparts. With this
modification, the doublon self-energy is given by
\begin{align}
\label{Eq:SigmaDIm}
\Sigma_d^{''}(\omega)&\!=\! z\myt^2C^{''}(\omega)\!-\!2z\myt^2\int_0^\omega \frac{d \omega '}{\pi} S^{''}(\omega ')\grfn_d^{''}(\omega\!-\!\omega ')\\
\Sigma_d^{'}(\omega)&\!=\! z\myt^2C^{'}(\omega)\!-\!2z\myt^2\int_{-\infty}^0 \frac{d\omega '}{\pi} S^{''}(\omega ')\grfn_d^{'}(\omega\!-\!\omega ') \nonumber\\
& 
\quad \quad \quad \quad +2z\myt^2\int_0^{\infty}\frac{d\omega '}{\pi}
\grfn_d^{''}(\omega ')S^{'}(\omega\!-\!\omega ')
\label{Eq:SigmaDRe}
\end{align}
where $\Sigma_d$, $C$, and $S$ are the retarded doublon self-energy,
Fermionic particle-particle and particle-hole propagators
respectively~\cite{ftnote_1}, and the primes $'$ and $''$ denote the
real and imaginary parts, respectively. The particle-particle and
particle-hole propagators are given by
\begin{align}
\label{Eq:SIm}
S^{''}(\omega)&\!=\!-\!\!\int_0^\omega \frac{d\omega '}{\pi}\grfn_f^{''}(\omega')\grfn_f^{''}(\omega'\!-\!\omega)\\
S^{'}(\omega)&\!=\!-\!\!\int_{-\infty}^0 \!\! \frac{d\omega '}{\pi}\grfn_f^{''}(\omega')[\grfn_f^{'}(\omega' \!-\! \omega)+\grfn_f^{'}(\omega'\!+\!\omega)]\\
C^{''}(\omega)&\!=\!-\!\!\int_0^\omega \frac{d\omega '}{\pi}\grfn_f^{''}(\omega')\grfn_f^{''}(\omega-\omega')\\
C^{'}(\omega)&\!=\int_{-\infty}^0 \frac{d\omega '}{\pi}\grfn_f^{''}(\omega')\grfn_f^{'}(\omega-\omega') \nonumber \\
\label{Eq:CRe}
&\quad\quad\quad\quad-\int_{0}^\infty \frac{d\omega '}{\pi}\grfn_f^{''}(\omega')\grfn_f^{'}(\omega'+\omega)]
\end{align}
where $\grfn_f(\omega)=\sum_\kk\grfn_f(\kk\omega)$ is the momentum
averaged Fermion Green function and the primes $'$ and $''$ denote the
real and imaginary parts. These equations, together with the equation
for the doublon Green function, Eq.~(\ref{Eq:dbl_gfn}), define a
system of self-consistent equations for the doublon self-energy. 

\begin{figure}
\includegraphics[scale=0.8]{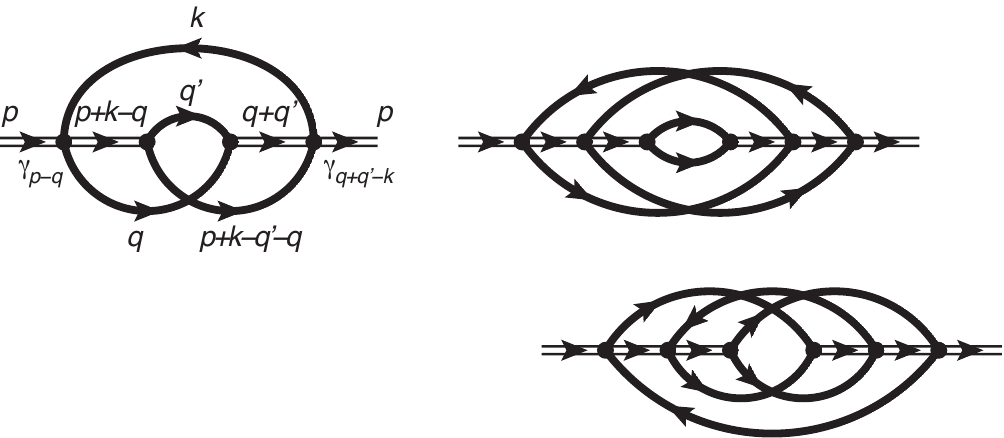}
\caption{Typical diagrams for the doublon Green function that are not
  accounted for in the resummation procedure as they contain crossing
  Fermion lines.  Here, double-lines stand for bare doublon
  propagators and thick single-lines the full (resummed) Fermion
  propagator. For reasons explained in the main text, these crossing
  diagrams do not contribute to the doublon decay, as the vertex
  factors are not paired and thus average to zero upon momentum
  integration. To see this, the momenta and a pair of vertices in the
  first, pretzel-like, diagram are labeled. Notice that the $\gamma$
  vertex factors have different momentum labels, these would have been
  identical for the case of a non-crossed diagram. }
\label{Fig:DoublonCross}
\end{figure}

In this section we have made two approximations: (1) we replaced the
momentum dependent vertex functions by momentum independent ones, and
(2) we have left out a large number of diagrams with crossing Fermion
lines (see Fig.~\ref{Fig:DoublonCross} for some typical examples).  To
verify these approximations, we have explicitly computed all diagrams
up to $6^{th}$ order in a Fermi Golden Rule calculation, which is free
of these approximations (see Appendix~\ref{appendix:FGR} for details).
We find that the decay rate computed via Fermi Golden Rule matches
very well with the resummation result. Further, within Fermi Golden
Rule calculation we empirically verify that the contribution of
crossed diagrams to the doublon self-energy is indeed
negligible. Intuitively, the reason for this seems to be that the
Fermion-doublon interaction vertex contains the factor
$\gamma_{\pp-\qq}+\gamma_\kk+\gamma_\qq$ which changes sign as we
sample momentum space.  The non-crossing diagrams involve squares of
this vertex function and do not change sign as we integrate over
momentum coordinates. On the other hand, the crossing diagrams involve
product of the vertices at different momenta and hence give a
negligible contribution upon integrating over momentum coordinates.

\subsection{Fermion Self-Energy}

\begin{figure}
\includegraphics[scale=0.8]{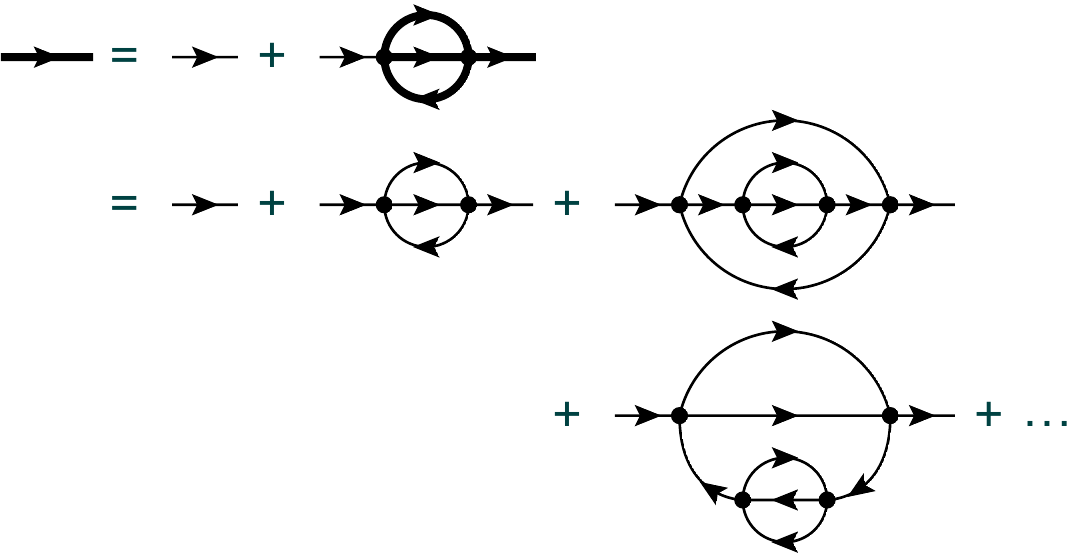}
\caption{Self-consistency equation for the Fermion propagator (top)
  and some typical diagrams that make up the full propagator
  (bottom). Thin lines indicate the bare propagator and thick
  lines the full (resummed) propagator.}
\label{Fig:FermionGF}
\end{figure}

We now come back to the question of evaluating the Fermion Green function 
\begin{equation}
\grfn_f(\kk)=\sum_k [\omega -\epsilon_\kk -\Sigma_f(\omega)]^{-1},
\label{Eq:GF}
\end{equation}
where $\epsilon_\kk=-\myt \, \gamma_\kk-\mu$ is the bare dispersion
and $\Sigma_f(\omega)$ is the Fermion self-energy that arises due to
interaction with other Fermions. To make progress, we begin by
considering the non-crossing approximation. As before, for the case of
the doublon self-energy, we are interested in the high frequency part
of the Green functions, and therefore (in the non-crossing
approximation) we are justified in replacing the vertices by their
momentum averaged counterparts as listed in first row of
Table~\ref{Table:Rules}, and then working with momentum averaged Green
functions and self-energies. In the non-crossing approximation, the
Fermion self-consistency equation is depicted diagrammatically in
Fig.~\ref{Fig:FermionGF}, where the thick Fermion lines represent
fully dressed Fermion Green functions that are being determined
self-consistently. The Fermion self-energy equations are given by
\begin{eqnarray}
\label{Eq:SigmaFRe}
\Sigma_f^{''}(\omega)& = &-2z\myt_1^2 \int_0^\omega \frac{d \omega '}{\pi} S^{''}(\omega')G^{''}(\omega -\omega') \\
\Sigma_f^{'}(\omega)& =& -2z\myt_1^2\int_{-\infty}^0 \frac{d\omega '}{\pi} S^{''}(\omega ')\grfn_f^{'}(\omega-\omega ') \nonumber \\
\label{Eq:SigmaFIm}
 & & \quad +2z\myt_1^2\int_0^{\infty}\frac{d\omega '}{\pi}\grfn_f^{''}(\omega ')S^{'}(\omega-\omega ').
\end{eqnarray}
Combining these self-energy equations with the definition of the Green
function Eq.~(\ref{Eq:GF}), we obtain a set of self-consistent
equations for the Fermion Green function.

\subsection{Corrections Due to Diagrams Left Out}
\begin{figure}
\includegraphics[scale=0.8]{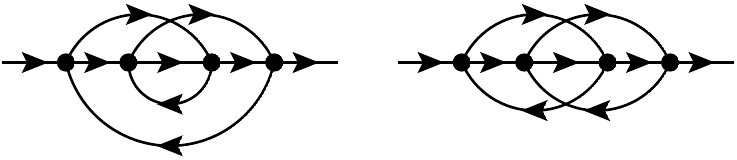}
\caption{Typical crossed fermion diagrams that are missed by the
  resummation method. These types of diagrams are expected to strongly
  contribute to the Fermion self-energy at high frequencies and
  therefore to the doublon decay rate.}
\label{Fig:FermionCross}
\end{figure}

\begin{figure}
\includegraphics[scale=0.8]{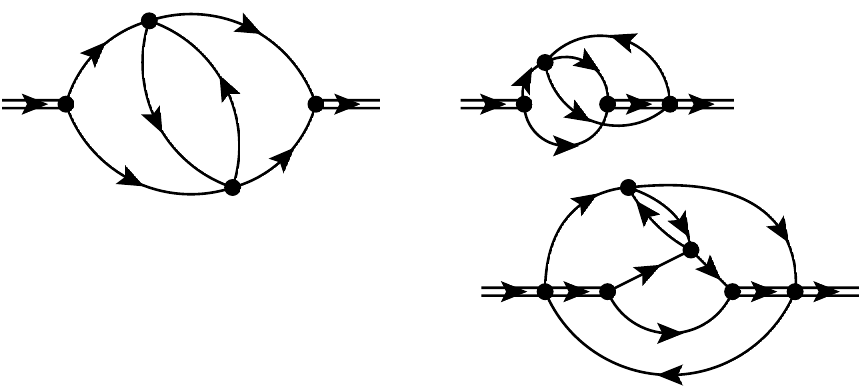}
\caption{Typical type III diagrams that are missed by the resummation
  method. As explained in the main text, these diagrams are expected
  to give some contribution to the doublon self-energy, but their
  effect is not taken into account.}
\label{Fig:TypeIII}
\end{figure}

In the resummation formalism we have missed three important classes of
diagrams: type I diagrams, which correspond to doublon self-energy
diagrams with crossing Fermion lines (examples depicted in
Fig.~\ref{Fig:DoublonCross}); type II diagrams, which are Fermion
self-energy diagrams with crossing Fermion lines (examples depicted in
Fig.~\ref{Fig:FermionCross}); and type III diagrams, which are doublon
self-energy diagrams which are left out and are neither type I nor
type II (examples depicted in Fig.~\ref{Fig:TypeIII}).

As mentioned earlier, we have empirically checked that type I diagrams
do not contribute to the doublon self-energy due to the lack of
pairing of the Fermion-doublon vertex factors. However, there are no
similar arguments for excluding type II or type III diagrams.  We
suppose that when a doublon emits a particle-hole pair, the particle
and hole are not coherent with each other, and therefore, we make the
approximation of dropping type III diagrams.  However, each Fermion in
the emitted pair still interacts with the Fermi sea, resulting in both
non-crossing Fermion self-energy diagrams, that have already been
taken care of, and type II diagrams which we shall try to estimate.

Since we cannot evaluate all the type II diagrams explicitly, we
proceed to approximate their effect on the Fermion self-energy in the
following way: 

(a) We assume that at a given frequency $\omega$, the leading
contribution to the imaginary part of self-energy
$\text{Im}\;\Sigma_f(\omega)$ comes from diagrams of a definite order
$n_0(\omega)$, as diagrams of lower order do not have enough
particle-hole pairs to absorb $\omega$ and diagrams of higher order
are suppressed by additional powers of $\myt/\omega$. We expect
$n_0(\omega)$ to scale linearly with $\omega$ as the main contribution
to the spectral function at $\omega$ comes from exciting $\sim
\omega/\epsilon_0$ particle-hole pairs, where $\epsilon_0$ is the
typical energy of particle-hole pairs.

\begin{figure}
\includegraphics[width=8cm]{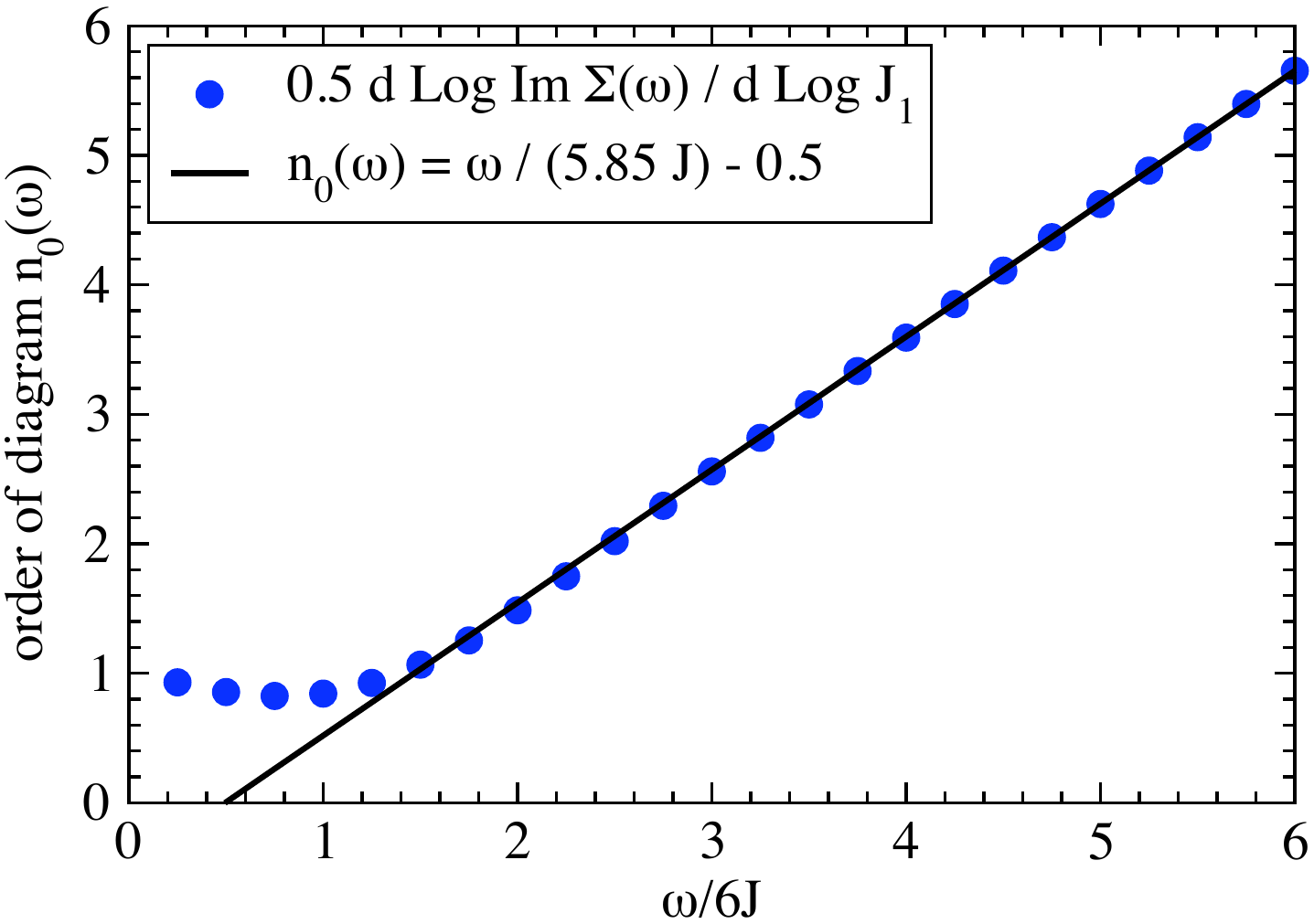}
\caption{Order with largest contribution to the Fermion self-energy
  $n_0(\omega)$ as a function of the frequency $\omega$. The solid line
  represents the best linear fit for the high frequency data. }
\label{Fig:OrderOfDiagram}
\end{figure}

(b) To determine $n_0(\omega)$, we keep the Fermion-Fermion vertex
energy scale $\myt_1$ as a free parameter, and calculate $n_0(\omega)$
from the logarithmic derivative
\begin{equation}
n_0(\omega)=\frac{1}{2}\left.\frac{d \log
\Sigma_f(\omega)}{d \log \myt_1}\right\vert_{\myt_1=\myt}.
\end{equation}
This relation is exact if only one order of diagrams contribute at
given energy; for the case of different orders contributing to
self-energy, this gives a number close to the order with leading
contribution. $n_0(\omega)$, obtained from the resummed self-energy,
is plotted in Fig.~(\ref{Fig:OrderOfDiagram}). The best fit for this graph
is $n_0(\omega)=\omega/(5.85J)-1/2$.

(c) We then compute the ratio of the total number of possible $n^\text{th}$
order Fermion self-energy diagrams to the number of $n^\text{th}$
order diagrams included in the resummation scheme, $R(n)$. $R(n)$ can
then be interpolated to form a function of the continuous variable
$n$. See Appendix~\ref{Appendix:Numerology} for details of computing
this ratio.

(d) In the final step, we rescale the imaginary part of the Fermion
self-energy by $R(n_0(\omega))$ to obtain a better approximation
including effects of missed diagrams
\begin{align}
\Sigma_f^{''}(\omega) \rightarrow \Sigma_f^{''}(\omega) R[n_0(\omega)].
\end{align}
Here, we are making an assumption: the amplitude of the
Fermion self-energy diagram only depends on its order in perturbation
theory and not on the details of the structure of the diagram.
Modulo the contribution of the type III diagrams, this approximation should 
overestimate the decay rate as the crossing diagrams usually contribute 
less than the non-crossing diagrams due to the momentum sums involved.
 
To complete the calculation of the doublon self-energy, we use the
Fermion Green function to construct the particle-particle and
particle-hole propagators Eqs.~(\ref{Eq:SIm}-\ref{Eq:CRe}), which
appear in the self-energy equations for the doublon
Eqs.~(\ref{Eq:SigmaDIm}, \ref{Eq:SigmaDRe}).

\section{Theoretical Results and Comparison with Experiments}

In this section we look at the theoretical results of the doublon lifetime
calculation and compare them with experimental results. We start by
summarizing the method of calculation, which will help in establishing
different approximation schemes. We then discuss the results from
different schemes and their comparison with experiments.

The calculation of the decay rate via the resummation technique has
two important steps. The first one is the evaluation of the
Fermion Green's functions which are used to compute the
particle-particle and particle-hole propagators. The second one is the
evaluation of the doublon self-energy, which uses these
propagators. As mentioned before, a non-crossing approximation for the
doublon self-energy yields good results. The crossing diagrams give
negligible contribution as the vertex functions which oscillate with
momenta kills the momentum averages. We also note that there is a set of 
doublon self-energy diagrams (the type III diagrams) which we neglect in 
our calculation.

Our approximations are then related to different ways of evaluating
the Fermion propagators. We consider three different approximations:
(i) Non-interacting Fermions; in this case we use the free Fermion
propagators with a band dispersion. One way of looking at this
approximation is to set $\myt_1=0$. (ii) Non-crossing approximation
for interacting Fermions; in this case we set $\myt_1=\myt$ but use
only non-crossing diagrams to evaluate the Fermion propagators. (iii)
Modified self-energy for interacting Fermions; in this case we modify
the self-energy of the interacting Fermions obtained by non-crossing
approximation to take into account Fermion self-energy diagrams missed
in the resummation. The modification procedure is detailed in the
previous Section.

\begin{figure}
\includegraphics[width=8cm]{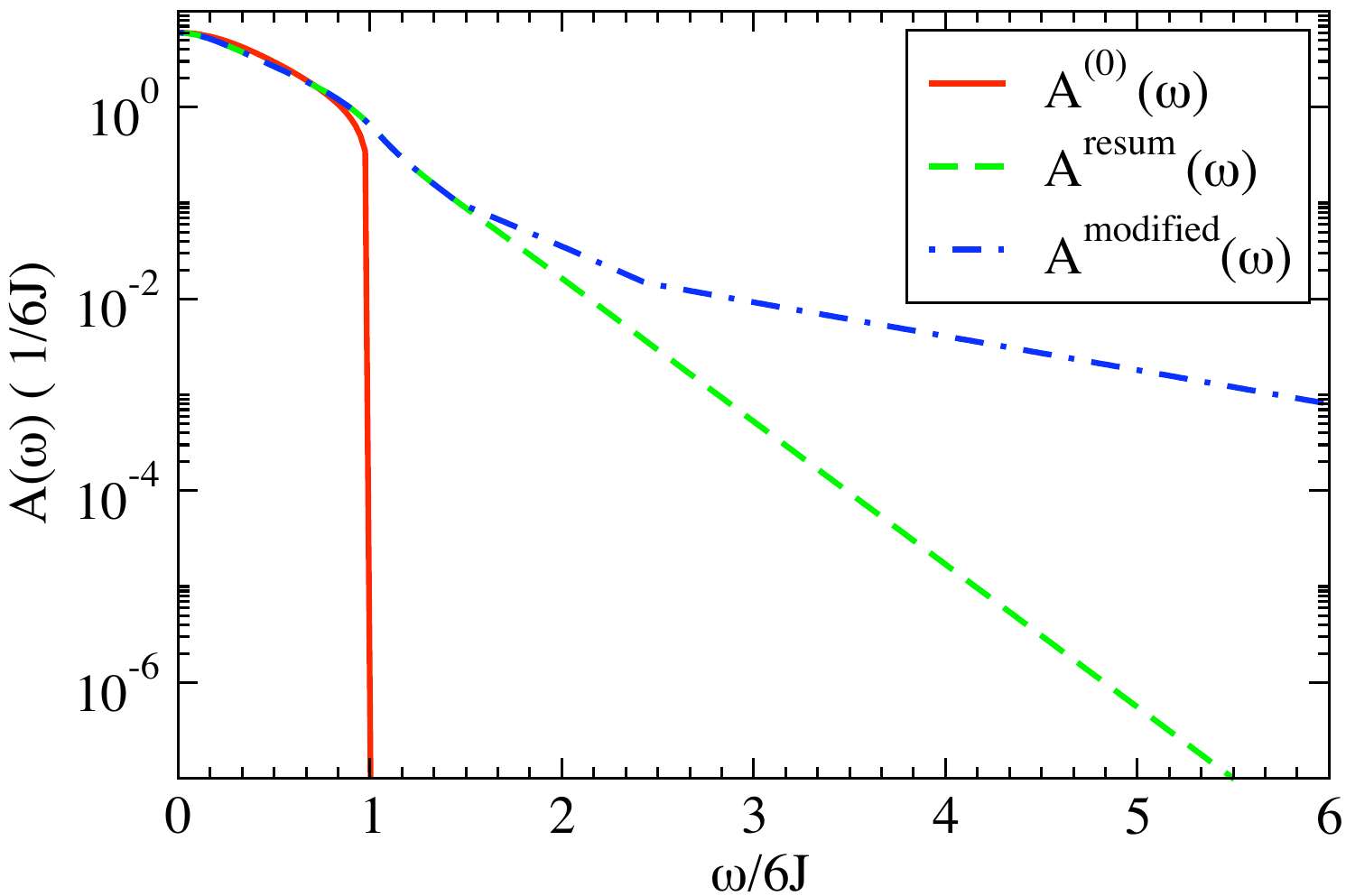}
\caption{Fermion spectral functions in different approximations: the free Fermion spectral function ($A^{(0)}(\omega)$); the projected
  Fermion spectral function obtained as the result of the resummation
  procedure ($A(\omega)$); the projected Fermion spectral function
  including corrections for missing diagrams in the resummation
  procedure (corrected $A(\omega)$). The linear slope at high energies on 
a semi-logarithmic scale shows the exponential transfer of spectral weight 
due to projection induced interactions.  }
\label{Fig:FermionSpectralFunction}
\end{figure}

We plot the spectral function of the Fermions,
$A(\omega)=-(1/\pi)\text{Im} {\cal G}_f(\omega)$, for the three
approximations in Fig.~\ref{Fig:FermionSpectralFunction}. In the
non-interacting case, this is simply the density of states in a cubic
lattice and the spectral weight is zero outside the band. In the
non-crossing approximation, we see that there is a transfer of
spectral weight from low energies to an exponential tail at high
energies, which reflects the fact that interaction induced by
projection leads to the possibility of creating a high energy Fermion,
which can reduce its energy by creating particle-hole pairs. This is
an important qualitative change that affects the physics of doublon
decay in a fundamental way. The interacting Fermion approximation
allows two distinct decay processes : (a) creation of several low
energy ($\omega \sim 2z\myt$) particle-hole pairs and (b) creation of
a high energy particle-hole pair which then decays into a shower of
low energy particle-hole pairs. The second process is forbidden for
non-interacting Fermions. Finally, in the modified self-energy
approximation, we include more processes to create particle-hole pairs
and hence there is a larger shift of spectral weight to higher
energies, as evidenced by the slower decay of the tail. This enhances
the importance of the (b) channel for decay.

\begin{figure}
\includegraphics[width=8cm]{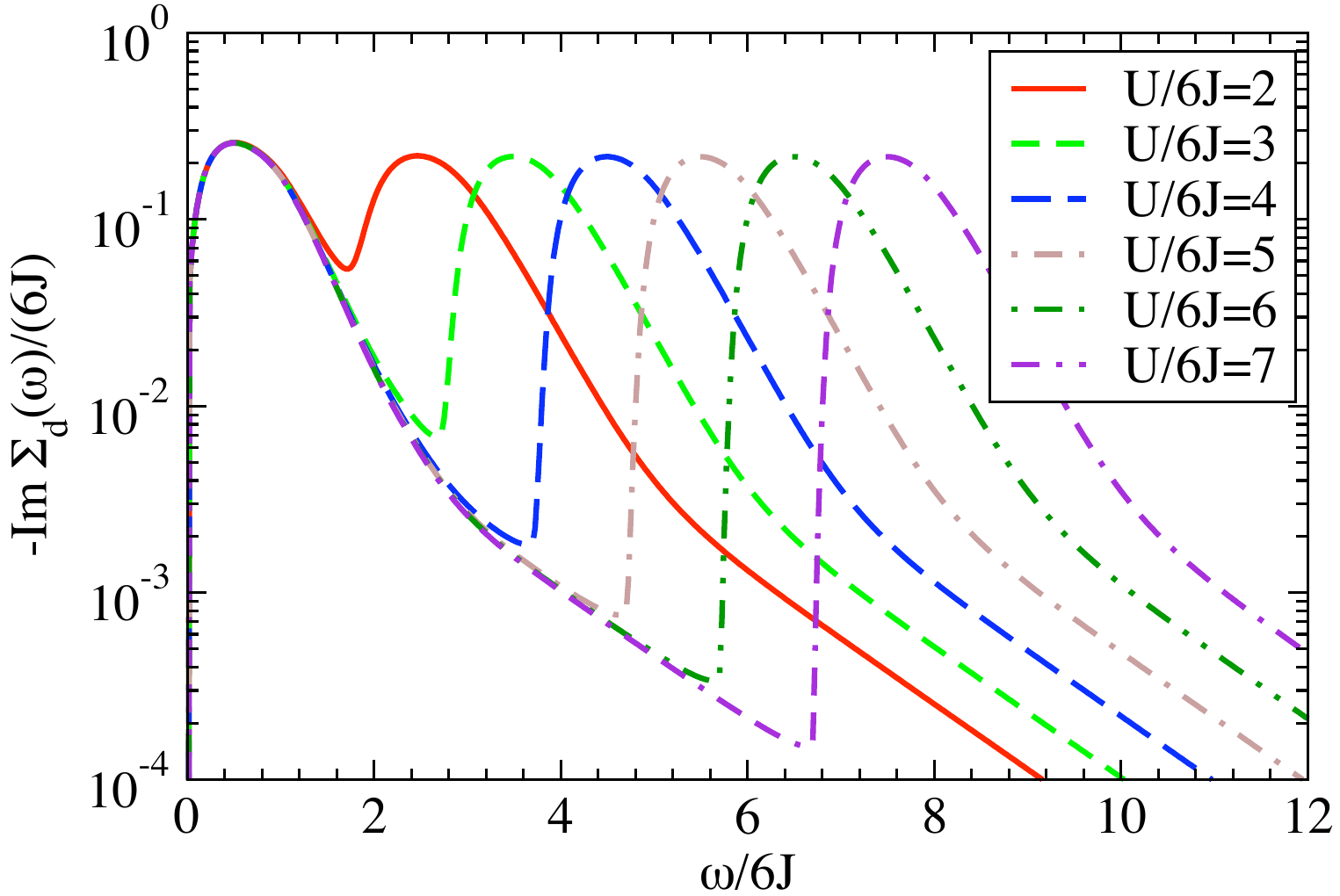}
\caption{Doublon self-energy (in the modified Fermion self-energy
  approximation) for various values of $U/6\myt$. }
\label{Fig:DoublonSelfEnergy}
\end{figure}
In the second step we use the Fermion propagator obtained in step
one to self-consistently compute the doublon self-energy. The
imaginary part of the doublon self-energy for various $U/6 \myt$
ratios is depicted in Fig.~\ref{Fig:DoublonSelfEnergy}.  The main
features are a pair of peaks, one occurring at small frequencies, and
another at high frequencies.  As there are no excitations in the Fermi
system in the initial state, for frequencies $\omega\leq U$ a nonzero
value of $\text{Im} \Sigma_d(\omega)$ corresponds directly to the rate of
doublon decay. At low frequencies, the doublon is far from its mass
shell and rapidly decays into a pair of particles. As the frequency
increases 
more and more particle-hole pairs are required to absorb the
doublon energy resulting in the exponential decrease in $\text{Im}
\Sigma_d(\omega)$. As $\omega$ surpasses $U$, a new contribution to the 
imaginary part of the doublon self-energy arises from processes where the 
doublon can scatter into a lower energy state
closer to the mass shell by releasing the excess energy in the form of
a few particle-hole excitations. This scattering process is responsible for
the high frequency peak in $\text{Im} \Sigma_d(\omega)$, that starts growing
at $\omega=U$. As we are interested in the decay of a doublon on the
mass shell, we read it from $\text{Im} \Sigma_d(U)$, which
corresponds to the smallest value of $\text{Im} \Sigma_d(\omega)$ between
the two peaks.

\begin{figure}
\includegraphics[width=8cm]{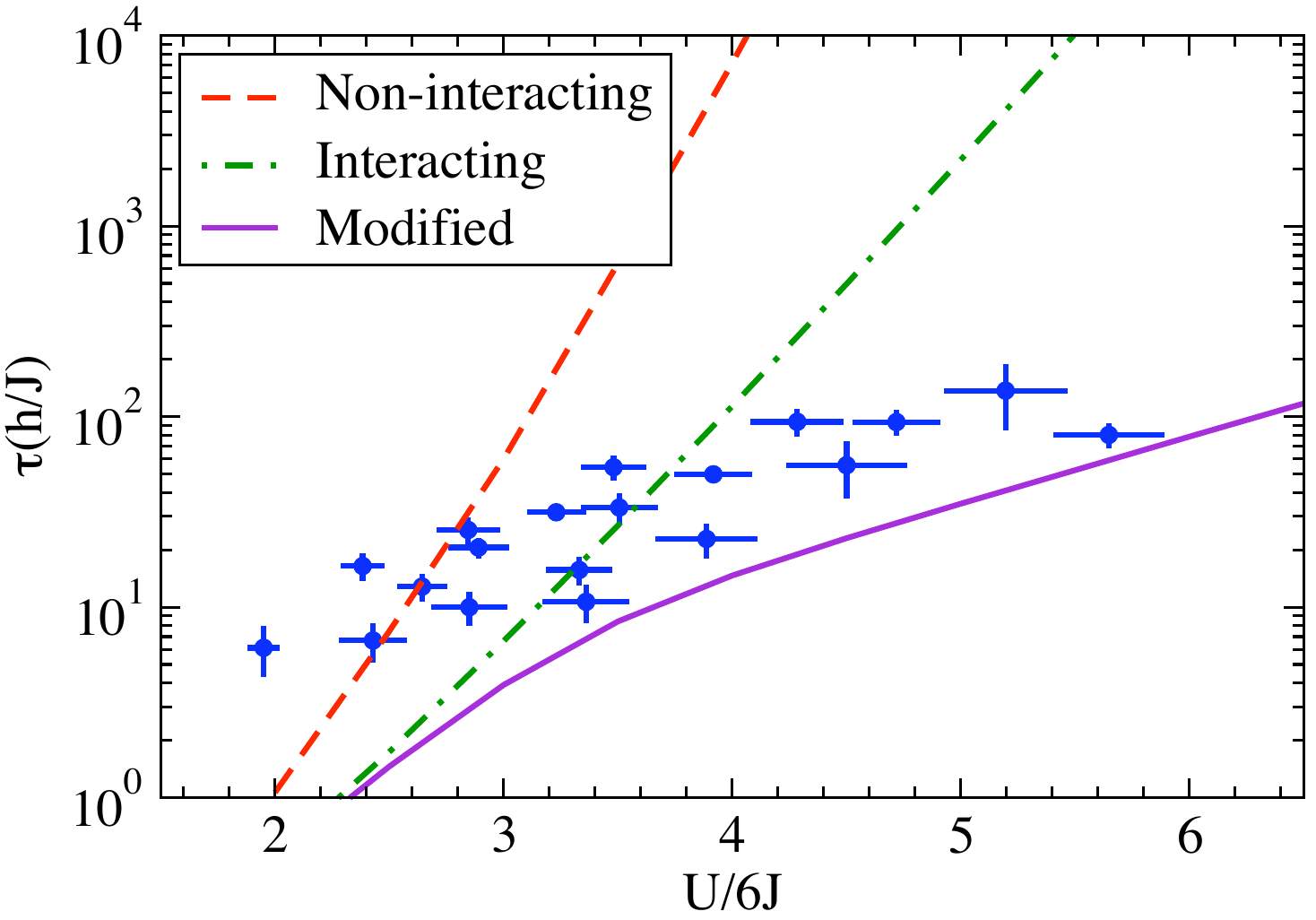}
\caption{Doublon decay time as a function of $U/6\myt$. The blue
  circles are the experimental data (cf.~Fig.~\ref{fig3:scalingdata}).
	The lines represent theoretical
  results from resummation with different levels of sophistication
  from non-interacting Fermions (red dashed line) to the non-crossing
  approximation with interacting Fermions (green dot-dashed line) to the
  modified self-energy approximation (purple solid line).}
\label{Fig:decay1}
\end{figure}
In Fig.~\ref{Fig:decay1}, we plot the experimentally obtained decay
time together with the theoretical estimates from the three different
approximations mentioned earlier. We proceed in the order of
sophistication, starting from the non-interacting Fermion case.  We
see that the decay time obtained with non-interacting Fermions
($\myt_1=0$) via resummation of doublon self-energy diagrams is much
longer than the experimentally obtained one. Setting $\myt_1=\myt$, and
using non-crossing diagrams for Fermion self-energy, we obtain a decay
time that is a closer match to the experimental data, but is still too
long. Next, we take care of the corrections to the imaginary part of
the Fermion self-energy from crossing diagrams and find a reasonable match
with experiments.

Finally, we want to comment on the remaining free parameters in our
calculation. The chemical potential of the Fermions, which determine
the hole density, is a free parameter, which can in principle be
determined from an equilibrium theory of a strongly interacting doped
Hubbard model. Since there is no consensus about the theory of the
doped Hubbard model, we prefer to keep it as a free parameter. We vary
it within the plausible range of $0.25\myt$ to $(-0.3 \myt)$ to see how
sensitive our results are to the choice of this parameter. The
dispersion in the lifetime is then plotted as the shaded region in
Fig.~\ref{fig3:scalingdata}. We see that we find good quantitative
agreement with the experiments in the slope of the lifetime curve,
i.e. for the co-efficient $\alpha$ in the exponent of the scaling function. The agreement in the prefactor $C$ is also fair, but this
quantity is sensitive to the choice of the free parameter in our
calculation.

\section{Concluding Remarks}

We have studied the decay of artificially created double occupancies
in the repulsive Fermi-Hubbard model in the presence of a
background compressible state. The situation is experimentally
realized by creating double occupancies and corresponding holes on top
of a half-filled system via optical lattice
modulation. Experimentally it is found that the decay time of the
doublons scales exponentially with $U/\myt$. We can understand the
observed scaling in terms of the fact that in order to decay the
doublon has to distribute its energy ($\sim U$) among $\sim U/\myt$
particle-hole excitations. We have developed a detailed theoretical
description of this process using diagrammatic resummation
techniques. Although the scaling form can be understood from a simple
energy conservation argument, we find that the co-efficient in the
exponent depends substantially on the strong interaction between the
background Fermions. After taking into account the effects of these
strong interactions, we find quantitatively fair agreement between
theory and experimental results.

The exponentially large lifetime of the doublons has serious
implications for use of cold atom systems to simulate the equilibrium
properties of the Hubbard model at large values of $U/\myt$.
Typically, in cold atom experiments, the strong interaction regime of
the Hubbard model is accessed by cooling the atoms in a weakly
interacting state and then tuning either the optical potential or the
magnetic field to change $U/\myt$. The lifetime of the doublons
constrains the maximum sweep rate of these Hamiltonian parameters
under which thermal equilibrium is maintained. As one goes towards
larger $U/\myt$, the sweep rates need to be exponentially slow to
maintain thermodynamic adiabaticity. Given intrinsic constraints like
lifetime of a sample, this would restrict the values of $U/\myt$ for
which the simulation of Hubbard model in thermal equilibrium can be
achieved.

However, this also opens up the possibility of studying
non-equilibrium dynamics of the Hubbard model, which may contain
interesting and new physics. In addition, the long lifetime of the
doublons also leads to the possibility of observing metastable states
with finite density of doublons. An intriguing scenario is observing
$\eta$ pairing of doublons and holes~\cite{etapair}.

Finally, we point out that similar phenomena may be relevant to the
issues of equilibration in the Bosonic Hubbard model. In a recent
paper~\cite{Chin2009} C. Chin's group observed the equilibration of
the density distribution of Bosonic atoms in a two dimensional optical
lattice after the lattice potential was ramped up. As the system
relaxed toward equilibrium, the center of the trap heated up, which
required the increase in the number of doublons. The slow relaxation
timescale observed in experiments may be a reflection of the ``dual''
problem to the one we discussed in this paper: slow rate of formation
of doublons from a a state containing only singly occupied sites and
holes.

\section{Acknowledgements}
We would like to thank A. Georges, A. Rosch, L. Glazman, and C. Chin
for useful discussions.  R.S., D.P., and E.D. acknowledge the
financial support of NSF, DARPA, MURI and CUA. E.A. acknowledges
support from BSF (E.D. and E.A.) and ISF. The experimental work was
supported by SNF, NAME-QUAM (EU) and SCALA (EU).

\appendix
\section{Model}
\label{appendix:Model}
In this appendix we derive the model we use to describe doublon decay
in the background of a projected Fermi sea. We begin with the
Fermi-Hubbard model
\begin{align}
H_\text{FH}=-\myt\sum_{\nbr\sigma}c^\dagger_{i,\sigma} c^{\phantom{\dagger}}_{j,\sigma}
+U\sum_i n_{i,\uparrow} n_{i,\downarrow},
\end{align}
where the first term describes the hopping of fermions and the second
term the on-site repulsive interaction.  We are interested in the case
$U \gg \myt$, where we expect doublons to be meta-stable particles.
Therefore, our goal is to decouple the doublon sector from the sector
of singles.  We do this by projecting out double occupancies from the
singles sector, and introducing doublon creation and annihilation
operators $d^\dagger_i$ and $d_i$ to take their place.  We, proceed in
two steps, first we use projection operators to separate the terms in
the Fermi-Hubbard Hamiltonian that preserve the number of doublons
from those that change it:
\begin{align}
H_\text{FH}=H_0+H_{+1}+H_{-1},
\end{align}
where $H_0$ preserves the number of doublons 
\begin{align}
\nonumber H_0&=-\myt\sum_{\nbr\sigma} (1-n_{i \bar{\sigma}}) \ca_{i \sigma} \cnn_{j \sigma}  (1-n_{j \bar{\sigma}})\\
\nonumber &-\myt\sum_{\nbr\sigma} (n_{i \bar{\sigma}}) \ca_{i \sigma} \cnn_{j \sigma}  (n_{j \bar{\sigma}})\\
&+U\sum_i n_{i \uparrow} n_{i \downarrow},
\end{align}
and $H_{\pm 1}$ increases/decreases it by one
\begin{align}
H_{+1}&=-\myt\sum_{\nbr\sigma} (n_{i \bar{\sigma}}) \ca_{i \sigma} \cnn_{j \sigma}  (1-n_{j\bar{\sigma}}),\\
H_{-1}&=-\myt\sum_{\nbr\sigma} (1-n_{i \bar{\sigma}}) \ca_{i \sigma} \cnn_{j \sigma}  (n_{j\bar{\sigma}}),
\end{align}
where $n_{i\sigma}=c^\dagger_{i \sigma} c^{\phantom{\dagger}}_{i
  \sigma}$ and $\bar{\sigma}$ indicates spin opposite to $\sigma$.  In
the second step, we replace double occupancies by the corresponding
doublon operators. Thus we have
\begin{align}
H_0=&-\myt\sum_{\nbr\sigma} (1-n_{i \bar{\sigma}})(1-n^d_{i}) \ca_{i \sigma} \cnn_{j \sigma}  (1-n^d_{j})(1-n_{j \bar{\sigma}}) \nonumber \\
&-\myt\sum_{\nbr\sigma} \da_i \dnn_j \cnn_{i\sigma} \ca_{j\sigma}
+U\sum_i n^d_{i},
\end{align}
and
\begin{align}
H_{+1}&=-\myt\sum_{\nbr\sigma} \sigma \da_j  (c_{j\bar{\sigma}} c_{i \sigma}) (1-n_{i\bar{\sigma}})\\
H_{-1}&=-\myt\sum_{\nbr\sigma} \sigma  (1-n_{i\bar{\sigma}}) (c^\dagger_{i \sigma} c^\dagger_{j\bar{\sigma}}) d_j,
\end{align}
where $n^d_i=\da_i \dnn_i$.  Thus far, we have obtained an expression
for the Fermi-Hubbard Hamiltonian that incorporates doublon
operators. This Hamiltonian was specifically derived in such a way as
to avoid creation of spurious states (e.g. a doublon and a single
fermion on the same site) by the use projection operators. As a
result, we do not need to supplement it with a constraint equation.

Now we can separate the terms in the Hamiltonian based on which sectors
they connect.  The Fermion-Fermion term arises from terms in $H_0$ that connect the projected sector and is given by
\begin{align}
H_f=-\myt\sum_{\nbr\sigma} (1-n_{i \bar{\sigma}}) \ca_{i \sigma} \cnn_{j \sigma}  (1-n_{j \bar{\sigma}}).
\end{align}
Likewise, the Doublon repulsion term also arises from $H_0$ and is given by
\begin{align}
H_d=U\sum_i n^d_i.
\end{align} 
The remaining terms connect the Fermion and Doublon sectors and are
\begin{align}
H_{fd}&=H_{+1}+H_{-1}+
\myt\sum_{\nbr\sigma}\\ 
&\left[(1-n_{i \bar{\sigma}})n^d_{j}+n^d_{i} (1-n_{j\bar{\sigma}}) + \da_j \dnn_i \right]
\ca_{i \sigma} \cnn_{j \sigma},
\end{align}
where we have dropped the term that is nonzero in the presence of a
pair of doublons as we are assuming that there is at most one
doublon. To complete the model, we drop terms that result in Feynman
vertices with more than two incoming and two outgoing propagators. We
have verified, numerically, that these diagrams do not significantly
contribute to the doublon decay rate.

\section{Checks on Approximations through Fermi Golden Rule Calculation}
\label{appendix:FGR}

In this appendix, we compute the doublon decay rate for the case of
non-interacting Fermions (i.e., we disregard $H_p$ part of the
Hamiltonian~(\ref{Eq:HF})). We treat $H_0=H_f^0+H_d$ as the base
Hamiltonian, and $H_{fd}$ as the perturbation Hamiltonian, and
evaluate the decay rate, via the Golden Rule, to very high order in
$H_{fd}$ using Monte Carlo integration. The objective of this appendix
is to test the approximations made in the resummation technique of
Section~\ref{Sec:diagramatics} on a simplified Hamiltonian. In
particular, we empirically verify that (1) we may ignore the crossing
diagrams in doublon self-energy and (2) we can use momentum averaged
Green functions to compute the decay rates. We begin by laying out the
formalism, and then list the results of Monte Carlo integration of
decay rates.

\subsection{Formalism}
Our goal is to compute the transition rate from the starting
configuration composed of a single doublon in a Fermi sea at finite
temperature to the final configuration composed of the initial
Fermi sea with the doublon converted into a pair of single particles
and a number of particle-hole excitations. The Fermi Golden rule
states that the decay rate is given by
\begin{align}
\Gamma(p) = \frac{2 \pi}{\hbar} \sum_f |\langle i | T | f \rangle|^2 \delta(E_i-E_f),
\end{align}
where the matrix element can be expressed in ordinary perturbation
theory via
\begin{align}
\langle f | T | i \rangle \!\! = \!\!\!\! \sum_{s_1, s_2, ...} \!\!\frac{\langle f | H_{fd} | s_{n-1} \rangle \langle s_{n-1} | H_{fd} | s_{n-2} \rangle ... \langle s_1 | H_{fd} | i \rangle}{(E_i-E_{s_1})(E_i-E_{s_2})...(E_i-E_{s_{n-1}})}.
\end{align}
Here, the sum goes over all intermediate states $s_i$, with energy
$E_{s_i}$, and $n$ is the order of perturbation theory.  In this
perturbation theory, the action of $H_{fd}$ (except for the final
matrix element $\langle f | H_{fd} | s_{n-1} \rangle$) is to create
particle-hole pairs. In principle, we may be able to connect the
initial state to the final state via other processes,
e.g.~doublon$\rightarrow$particle-particle$\rightarrow$doublon,
however, these process lead to decay at higher order in perturbation
theory, and thus we ignore them.

We label the initial state by the momentum of the doublon $p$:
\begin{align}
|i\rangle=|\cdot;p\rangle=d^\dagger_p |\text{FS}\rangle.
\end{align}
Likewise, we label the final state via a set of momenta for the up (down)
spin particles $k_{i\uparrow(\downarrow)}$ and the up (down) spin holes
$q_{i\uparrow(\downarrow)}$:
\begin{align}
|f\rangle &= \left|
\begin{array}{cc}
k_{1,\uparrow}...k_{n_\uparrow+1,\uparrow},&k_{1,\downarrow}...k_{n_\downarrow+1,\downarrow},\\
q_{1,\uparrow}...q_{n_\uparrow,\uparrow},&q_{1,\downarrow}...q_{n_\downarrow,\downarrow} 
\end{array};\cdot\right\rangle\\
&=c^\dagger_{k_{n_\uparrow+1,\uparrow}} c^\dagger_{k_{n_\downarrow+1,\downarrow}}
\left(c^\dagger_{k_{n_\downarrow,\downarrow}} c_{q_{n_\downarrow,\downarrow}}\right)
...
\left(c^\dagger_{k_{1,\downarrow}} c_{q_{1,\downarrow}}\right) \times \nonumber \\
&\quad\quad\quad
\times \left(c^\dagger_{k_{n_\uparrow,\uparrow}} c_{q_{n_\uparrow,\uparrow}}\right)
...
\left(c^\dagger_{k_{1,\uparrow}} c_{q_{1,\uparrow}}\right) |\text{FS}\rangle,
\label{Eq:fs}
\end{align}
where $n_{\uparrow\,(\downarrow)}$ counts the number of spin up (down)
particle-hole pairs created ($n_\uparrow+n_\downarrow+1=n$). 

The intermediate states are composed of a doublon and $1,2,3,...,n-1$
fermion-hole pairs. Using $H_{fd}$, we can write the matrix element as
\begin{widetext}
\begin{align}
  \langle f | T | i \rangle &= \left\langle 
\begin{array}{cc}
k_{1,\uparrow}...k_{n_\uparrow+1,\uparrow},&k_{1,\downarrow}...k_{n_\downarrow+1,\downarrow},\\
q_{1,\uparrow}...q_{n_\uparrow,\uparrow},&q_{1,\downarrow}...q_{n_\downarrow,\downarrow} 
\end{array};\cdot
\right| T \left| \cdot ;p \right\rangle \\
&= \sum_{\text{permutations}} 
\text{sig}(\text{perm})
  \frac{\langle
    f | H_1 |
    (\tilde{\sigma}_1,\tilde{k}_1,\tilde{q}_1),...,(\tilde{\sigma}_{n-1},\tilde{k}_{n-1},\tilde{q}_{n-1});
    \tilde{p}_{n-1} \rangle ... 
    \langle (\tilde{\sigma}_1,\tilde{k}_1, \tilde{q}_1); \tilde{p}_1 |
    H_1 | p \rangle}{(\xi_{\tilde{k}_1}+...+\xi_{\tilde{k}_{n-1}} -
    \xi_{\tilde{q}_1} - ... - \xi_{\tilde{q}_{n-1}})
    ... (\xi_{\tilde{k}_1} + \xi_{\tilde{k}_2} - \xi_{\tilde{q}_1} -
    \xi_{\tilde{q}_2})(\xi_{\tilde{k}_1} - \xi_{\tilde{q}_1})},
\label{Eq:ME}
\end{align}
\end{widetext}
where $\tilde{k}_i$, $\tilde{q}_j$, $\tilde{p}_v$ stand for the
particle, hole, and doublon momenta, respectively, and
$\tilde{\sigma}_i$ indicates the spin of the $i$-th particle-hole
pair. The sum runs over all intermediate states that lead to the final
state $|f\rangle$. That is, we must sum over all permutations of
assigned values to $(\tilde{\sigma}_i, \tilde{k}_i, \tilde{q}_j)$ from
the list $\{k_{1,\uparrow},...,k_{n_\uparrow+1}\}$,
$\{k_{1,\downarrow},...,k_{n_\downarrow+1}\}$,
$\{q_{1,\uparrow},...,q_{n_\uparrow}\}$,
$\{q_{1,\downarrow},...,q_{n_\downarrow}\}$.  Within this labeling
scheme, the doublon momenta in the intermediate states $\tilde{p}_v$,
and the hole momentum in the final state, are chosen automatically by
momentum conservation. We take care of the Fermionic anti-commutation
relations with $\text{sig}(\text{perm})$, which stands for the
signature of the permutation, and is $\pm 1$ for even/odd permutations
of momenta.

\begin{figure}
  \includegraphics[width=8cm]{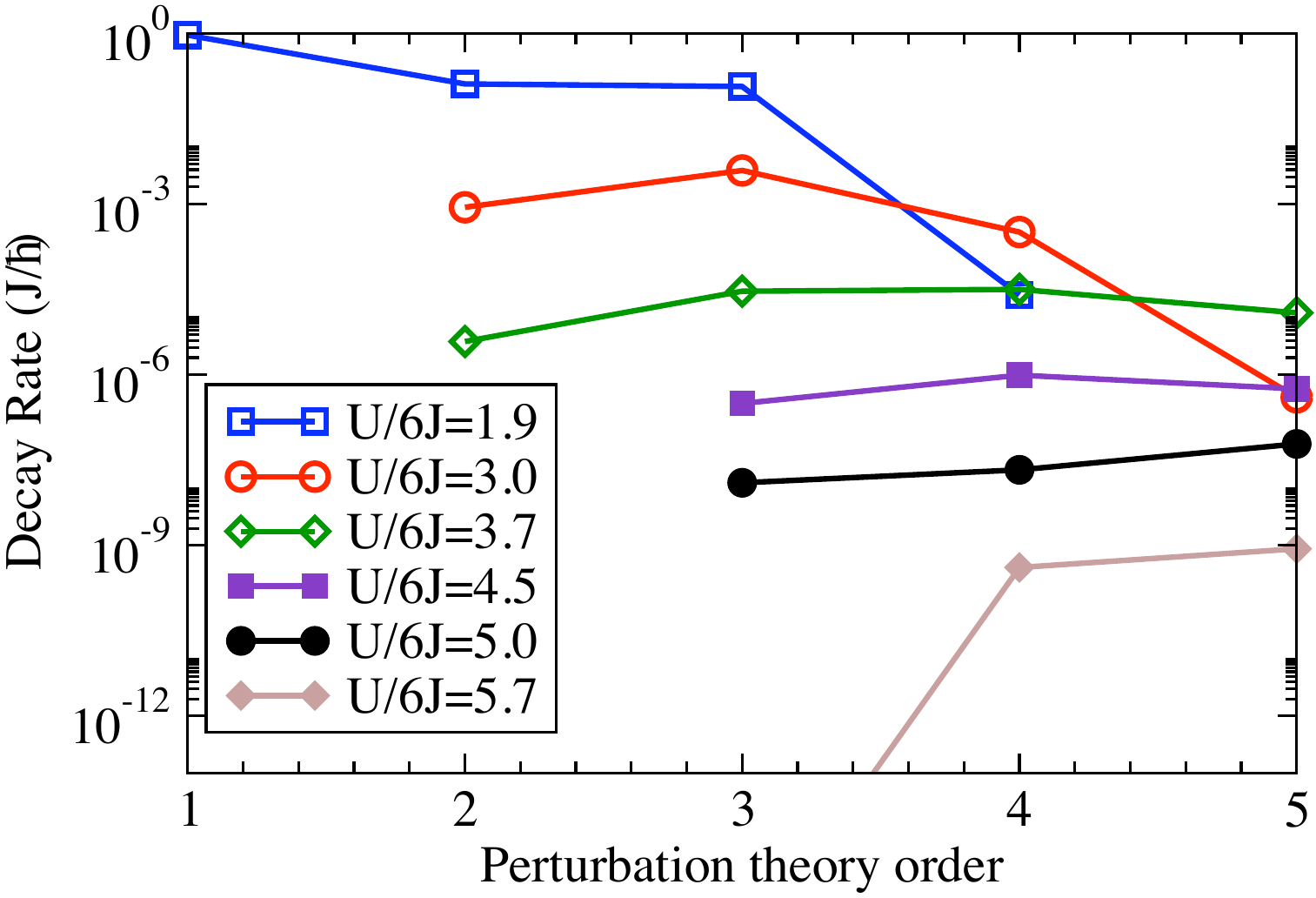}
  \caption{Decay rate as a function of the order of the perturbation
    theory computed using Fermi Golden rule. Largest decay rate
    corresponds to most important order.}
  \label{Fig:tau_vs_n}
\end{figure}

To obtain the decay rate, we trace over the final states, order by
order in perturbation theory,
\begin{align}
\Gamma(p)=\sum_{n=0}^\infty \Gamma^{n}(p).
\end{align}
At each order we trace over the number of up- and down-spin
particle-hole pairs, and the corresponding momenta of particles and
holes that make up the final state.  The decay rate at $n$-th order
is the given by the expression
\begin{widetext}
\begin{align}
  \Gamma^{n}(p) = \frac{2 \pi}{\hbar} \sum_{n_\uparrow+n_\downarrow+1=n} 
  \int  \frac{
  \left[\dbar k_{1,\uparrow} ... \dbar k_{n_\uparrow+1,\uparrow}\right]\,
  \left[\dbar k_{1,\downarrow} ... \dbar k_{n_\downarrow+1,\downarrow}\right]\,
  \left[\dbar q_{1,\uparrow} ... \dbar q_{n_\uparrow,\uparrow}\right]\,
  \left[\dbar q_{1,\downarrow} ... \dbar q_{n_\downarrow,\downarrow}\right]}{(n_\uparrow+1)! (n_\downarrow+1)! (n_\uparrow)! (n_\downarrow)!} \nonumber \\
 \quad\quad\quad  \quad\quad\quad \quad\quad\quad
 \delta(U-E_f) \, \delta\left(p-\sum k+\sum q \right)
\left| \left\langle 
\begin{array}{cc}
k_{1,\uparrow}...k_{n_\uparrow+1,\uparrow},&k_{1,\downarrow}...k_{n_\downarrow+1,\downarrow},\\
q_{1,\uparrow}...q_{n_\uparrow,\uparrow},&q_{1,\downarrow}...q_{n_\downarrow,\downarrow} 
\end{array};\cdot
\right| T \left| \cdot ;p \right\rangle
\right|^2,
\label{Eq:FGRk}
\end{align}
\end{widetext}
where $\dbar k$ stands for $f(k)\, d^3k/(2 \pi)^3$, $\dbar q$ for
$(1-f(q))\, d^3q/(2 \pi)^3$, and $f(k)$ is the Fermi function. The
denominator in the integral takes care of the fact that interchanging a
pair of momentum labels does not change the final state, $E_f=
\xi(k_{1,\uparrow})+...+\xi(k_{n_\uparrow+1,\uparrow})
+\xi(k_{1,\downarrow})+...+\xi(k_{n_\downarrow+1,\downarrow})
-\xi(q_{1,\uparrow})-...-\xi(q_{n_\uparrow,\uparrow})
-\xi(q_{1,\downarrow})-...-\xi(q_{n_\downarrow,\downarrow}) )$ is the
final state energy, and the second $\delta$ function takes care of
momentum conservation.
 \begin{figure}
   \includegraphics[width=8cm]{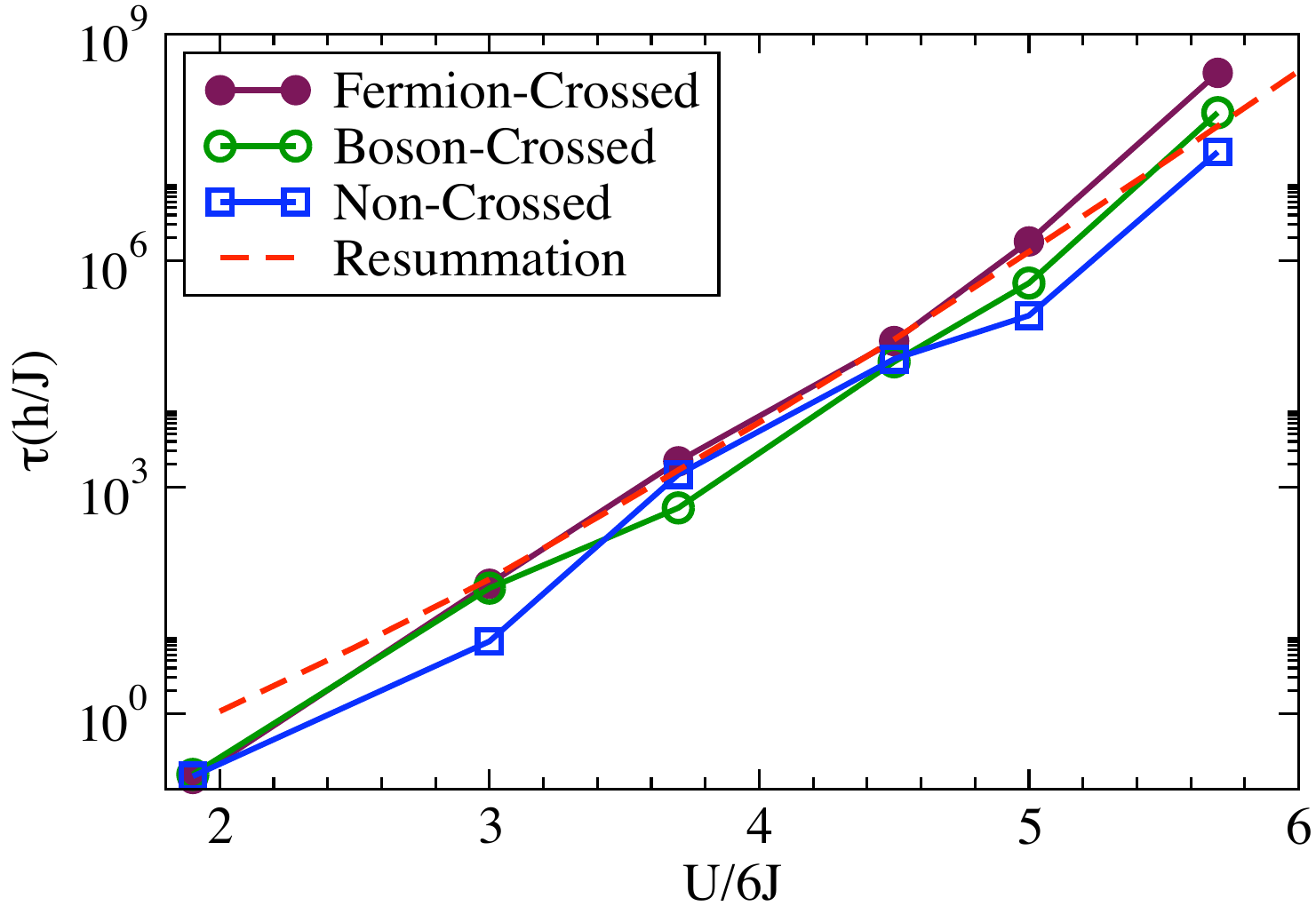}
   \caption{Comparison of the resummation method and various Golden Rule 
     approximations for calculating the dependence of the Doublon decay time 
     on the interaction strength $U/6J$ (with non-interacting
     Fermions).  
   }
  \label{Fig:FGRCOMP}
\end{figure}

We explicitly evaluate the $3^{2n}$ dimensional integral in
Eq.~(\ref{Eq:FGRk}) numerically via Monte Carlo integration.  To
perform this integration, we replace the $\delta$ function of energy,
which defines a hypersurface in momentum space -- a volume of of
measure zero, by the top hat function. We also use important sampling
to speed up integration by biasing our selection so that we pick
particle-hole pairs with holes in the Fermi sea and particles outside
of it. The main numerical constraint on the speed of integration
comes from evaluating the $(n_\uparrow+1)!(n_\downarrow+1)!
n_\uparrow! n_\downarrow!$ permutations over the 
intermediate states, which becomes rather expansive for
$n>6$.

\subsection{Results}

We begin by verifying that the perturbation theory in $H_{fd}$ does
indeed converge. That is, for fixed $U/6J$, does $\Gamma^{(n)}(p)$
decrease sufficiently fast as $n$ increases? We know that for $n
\lesssim U/12J$, $\Gamma^{(n)}(p)=0$, as not enough particle-hole
pairs are formed to carry away the energy of a doublon.  When $n \sim
U/12J$, in order to satisfy energy conservation, particles created in
the decay must have momentum in vicinity of the band maximum near
$(\pi,\pi,\pi)$ and holes in the vicinity of the band minimum at
$(0,0,0)$. Therefore, for $n \sim U/12J$ the volume of the momentum
space being integrated is very small, but this volume increases
quickly as $n$ grows. As a result, we expect that the
$\Gamma^{(n)}(p)$ will increase with $n$ for small $n$. On the other
hand, at high orders the decay rate is suppressed by a high powers of
the small parameter $J/U$. Thus, we expect $\Gamma^{(n)}(p)$ to have a
maximum for some intermediate value of $n$ close to, but somewhat
larger than $U/12J$.

In Fig.~\ref{Fig:tau_vs_n} we plot $\Gamma^{(n)}(p)$ as a function of
$n$ for various values of $U/6J$. In all cases, computations have been
performed at $T=0$ and $\mu=0$ (corresponding to one particle per two
sites). As expected, in all cases, we see a clear peak in
$\Gamma^{(n)}(p)$ at $n \sim U/12J + 2$.

Having verified the convergence of the high order perturbation
expansion, we move on to empirically verify whether we can ignore
crossing diagrams, at least for the case of free Fermions.  In order
to perform this comparison we compute the total decay rate as a
function of $U/6t$ using both Monte Carlo integration of
Eq.~(\ref{Eq:FGRk}) (incorporates all possible diagrams), as well as
the resummation of the non-crossing diagrams given by
Eq.~(\ref{Eq:SigmaDRe}) with bare Fermion Green functions used to
compute $C(\omega)$ and $S(\omega)$. We perform two additional tests
using Monte Carlo integration: (1) We calculate the decay rate with
Bosonic instead of Fermionic signs for closed Fermion loops; (2) We
keep only the diagonal terms, i.e. we replace $|\sigma_\text{perm}
... |^2\rightarrow \sigma_\text{perm} | ... |^2$, which corresponds to
the order-by-order summation of non-crossing diagrams, but without
momentum averaging of the resummation approach.  The results of these
four types of calculations are plotted in Fig.~\ref{Fig:FGRCOMP}, for
$T=0$ and $\mu=0$.  There is very good agreement between all four
cases, confirming that crossing diagrams may indeed be dropped as
explained in subsection~\ref{Subsection:DoublonSelfEnergy}.
\begin{figure}[h!]
\includegraphics[scale=0.8]{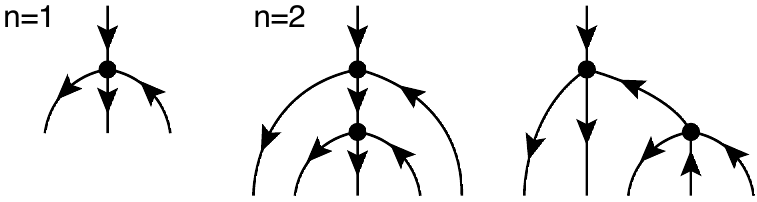}
\caption{All distinct tree diagrams with one vertex (left) and two
  vertices (right).}
\label{Fig:Tree}
\end{figure}

\section{Diagram Counting}
\label{Appendix:Numerology}
\begin{figure}
\includegraphics[width=8cm]{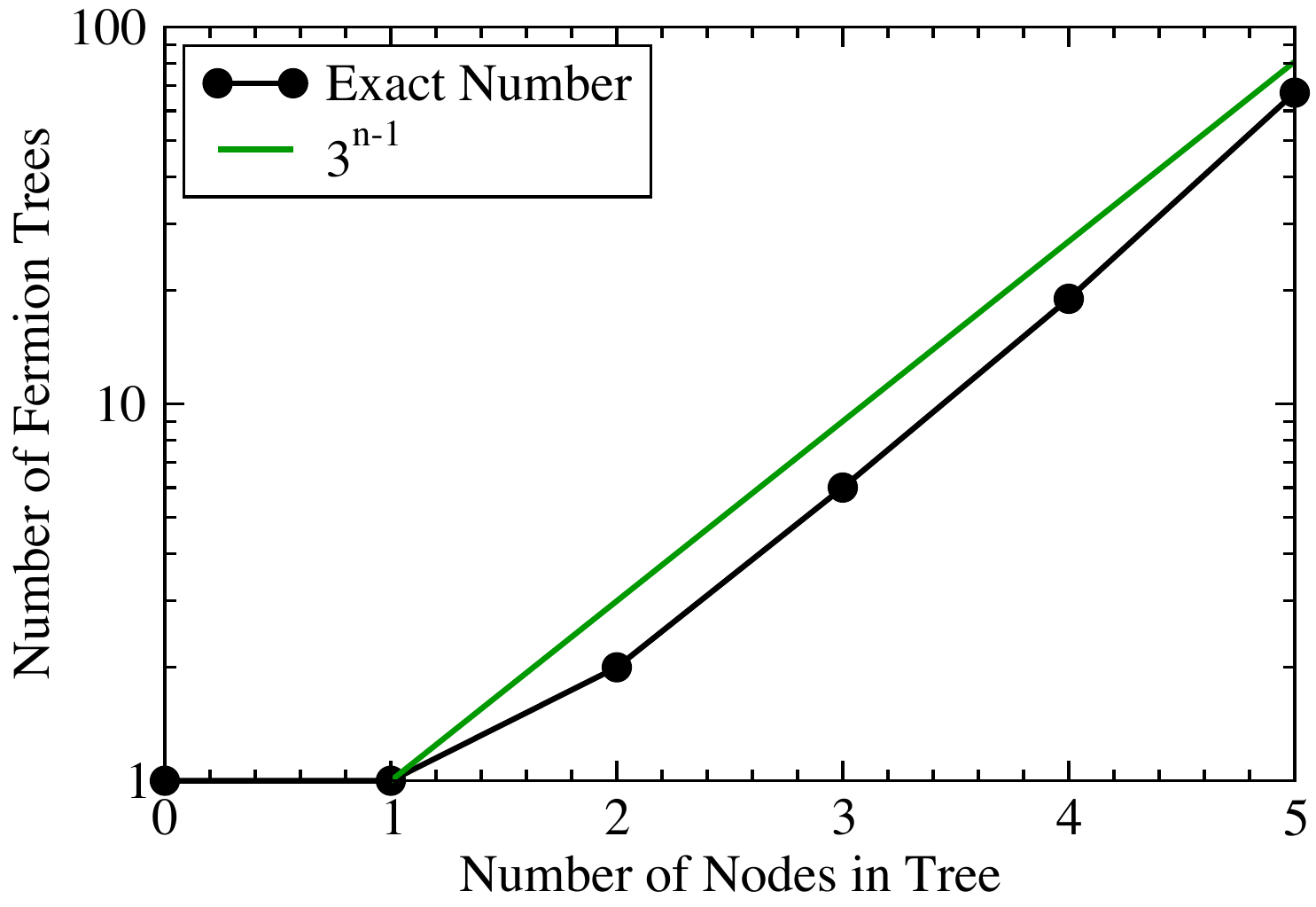}
\caption{Dependence of the number of distinct tree diagrams on the number of nodes in the tree. }
\label{Fig:nTree}
\end{figure}
In this appendix we describe the procedure for counting the total
number of distinct, spin-labeled Fermion self-energy diagrams at a
given order $Q_\text{all} (n)$ and the number of non-crossed
spin-labeled Fermion self-energy diagrams $Q_\text{nc} (n)$. We remind
the reader that $Q_\text{all} (n)$ and $Q_\text{nc} (n)$ correspond to
diagrams with $2 n$ vertices. For high $\omega$,
$\Sigma_f^{''}(\omega)$ is dominated by diagrams with maximal number
of particle and hole lines in the middle, as these maximize the energy
that is being transferred to the particle-hole pairs being created. In
fact, the range in $\omega$ over which $\Sigma_f^{''}(\omega)$ is
nonzero is proportional to the number of particle- and hole-lines in
the middle of the diagram. Therefore, to simplify the counting, we
only count diagrams that have the maximal number ($2n +1$) of
particle- and hole-lines going across the middle of the diagram.

To count the number of diagrams at given $n$, we first construct all
distinct tree diagrams (without spin labels) that have a single
particle going in, $n+1$ particles and $n$ holes going out and $n$
vertices of the type given in first row of Table~\ref{Table:Rules}. In
Fig.~\ref{Fig:Tree}, we show all such tree diagrams for $n=1$ and
$n=2$. In Fig.~\ref{Fig:nTree} we show how the number of distinct
trees scales with $n$.
\begin{figure}
\vspace{0.5cm}
\includegraphics[width=8cm]{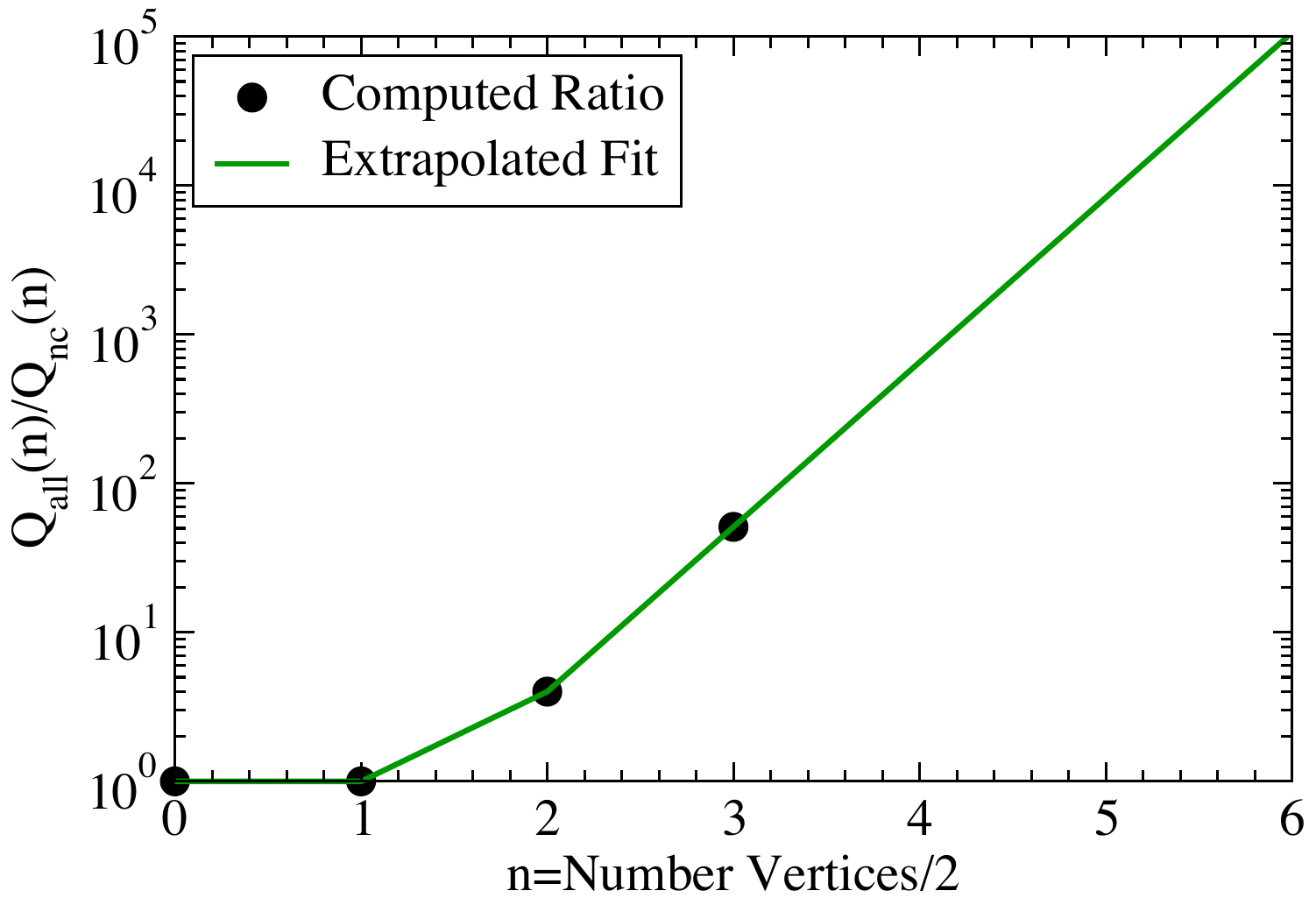}
\caption{Correction ratio as a function of the order of the diagram.}
\label{Fig:correctionRatio}
\end{figure}
Next, we construct the set of all the possible self-energy diagrams by
taking a pair of tree diagrams, reversing all the arrows in one of
them, and gluing them together. When we count the total number of
diagrams, we glue together particle-particle lines and hole-hole lines
in all pairs of trees at the given order, in all possible ways. On the
other hand, when counting the number of diagrams produced by the
non-crossing approximation, we only glue together trees with their
mirror image. Finally, we spin label the resulting diagrams, and
remove all duplicate diagrams, to obtain $Q_\text{all} (n)$ and
$Q_\text{nc} (n)$.

We assume that the ratio $Q_\text{all} (n)/Q_\text{nc} (n)$ scales
like $\sim e^{\alpha n}$. We use this assumption to extrapolate the
ratio for non-integer values of $n$ and for large values of $n>4$. We
plot the ratio of $Q_\text{all} (n)/Q_\text{nc} (n)$, along with the
extrapolated curve that we use in rescaling the Fermion self-energy,
in Fig.~\ref{Fig:correctionRatio}.


\begin{thebibliography}{99}
\bibitem{ftnote_1} The pair propagators are momentum integrated objects e.g. $S(\omega)=\sum_{\qq} S(\qq,\omega)$ and so on
\end{thebibliography}
\end{document}